# Semiclassical Quantization of Billiards with Mixed Boundary Conditions

M. Sieber[1,2], H. Primack[1], U. Smilansky[1], I. Ussishkin[1], H. Schanz[3]

[1] Department of Physics of Complex Systems, The Weizmann Institute of Science, Rehovot 76 100, Israel

[2] Division de Physique Théorique, Institut de Physique Nucléaire, 91 406 Orsay CEDEX, France

[3] Institut für Physik, Humboldt - Universität, 10 099 Berlin, Germany

February 19, 1995

**Abstract**

The semiclassical theory for billiards with mixed boundary conditions is developed and explicit expressions for the smooth and the oscillatory parts of the spectral density are derived. The parametric dependence of the spectrum on the boundary condition is shown to be a very useful diagnostic tool in the semiclassical analysis of the spectrum of billiards. It is also used to check in detail some recently proposed parametric spectral statistics. The methods are illustrated in the analysis of the spectrum of the Sinai billiard and its parametric dependence on the boundary condition on the dispersing arc.

## 1 Introduction and Statement of the Main Results

A classical billiard is completely defined once its boundary $\Sigma$ is prescribed. Then, the requirement that the particles (rays) reflect specularly determines the dynamics. To address the quantum (wave) analogue, it is necessary to solve the Schrödinger equation, which reduces for billiards to the Helmholtz equation

$$(\Delta + k^2)\psi(\vec{r}) = 0 \,. \tag{1}$$

$k$ is the wave number under consideration and we use natural units where $\hbar = 2m = 1$ and $E = k^2$.

In wave dynamics, one needs to supply an additional piece of information, that is, a condition which the wave function has to satisfy on the boundary. In quantum mechanical applications, one often considers the boundary as high potential wall whose height tends to infinity and then the boundary condition is the Dirichlet condition

$$\psi(\vec{r}) = 0 \,, \ \vec{r} \in \Sigma \,. \tag{2}$$

Another common boundary condition is the Neumann boundary condition

$$\partial_{\hat{n}}\psi(\vec{r}) = 0 \,, \ \vec{r} \in \Sigma \,, \tag{3}$$

where $\partial_{\hat{n}}$ stands for the normal derivative, with the normal pointing *outside*. This boundary condition appears most naturally for the pressure field in acoustics. One can generalize the above boundary conditions by requiring

$$\kappa(\vec{r})\psi(\vec{r}) + \partial_{\hat{n}}\psi(\vec{r}) = 0 \ , \ \vec{r} \in \Sigma \ . \tag{4}$$

This form leaves the problem self-adjoint when $\kappa$ is real. The Dirichlet boundary condition is recovered by setting $\kappa(\vec{r})^{-1} = 0$. We shall restrict the discussion to non-negative $\kappa(\vec{r})$ functions. If $\kappa(\vec{r})$ is negative, the Helmholtz equation can have solutions for imaginary values of the wave number $k$. Such solutions are unphysical, since they correspond to steady states despite of a dispersion of energy in the medium that takes place when the imaginary part of $k$ is different from zero [1]. Furthermore, we shall not deal with the most general positive $\kappa(\vec{r})$ but restrict our attention to piecewise constant functions on the boundary. In some applications it is convenient to express eq. (4) in terms of a positive parameter $b$ and a (piecewise constant) mixing angle $\alpha$ so that the boundary condition reads

$$b \cos\alpha \, \psi(\vec{r}) + \sin\alpha \, \partial_{\hat{n}}\psi(\vec{r}) = 0 \ , \ \vec{r} \in \Sigma \ . \tag{5}$$

This form interpolates conveniently between the Dirichlet and the Neumann boundary conditions, in terms of $\alpha$ which is limited to the interval $0 \leq \alpha \leq \frac{\pi}{2}$.

At this point we would like to make the following observation. In the semiclassical domain, the dominant term in the boundary condition (4) is $\partial_{\hat{n}}\psi(\vec{r})$ which is of order $k$. Hence, for a fixed $\kappa$ and in the semiclassical limit ($k \to \infty$), the spectrum will always tend to the Neumann case. A more proper definition of the semiclassical limit is one, that allows also in the limit $k \to \infty$ an interpolation between the Dirichlet and Neumann boundary conditions. This can be achieved by considering $k$ and $\kappa/k$ as independent parameters when performing this limit. To emphasize this point, we shall always quote the results as functions of these two parameters.

Mixed boundary conditions are not encountered in physical applications as often as Dirichlet and Neumann boundary conditions. A perturbative treatment of the Helmholtz equation with respect to changes in boundary conditions can be found in [2], but the semiclassical quantization of billiards with mixed boundary conditions was rarely discussed previously. The only exception we found was the work of Balian and Bloch, who encountered these boundary conditions in a nuclear physics context [3].

Our interest in the mixed boundary conditions stems from the observation that one can use the additional freedom provided by the function $\kappa(\vec{r})$ as a powerful tool in the analysis of various aspects of the semiclassical quantization of billiards. To explain this rather non-conventional approach, we shall have to start by quoting the results of the semiclassical derivation for two–dimensional billiard systems which is presented in the next two sections.

We consider the spectral density in terms of the wave number $k$. For positive $k$ and $k_n(\kappa) = +\sqrt{E_n(\kappa)}$ it is given by

$$d(k;\kappa) = \sum_{n=1}^{\infty} \delta\left(k - k_n(\kappa)\right) = \bar{d}(k;\kappa) + d_{osc}(k;\kappa) \ , \tag{6}$$

which depends parametrically on the boundary conditions. The semiclassical theory uses separate methods to evaluate the smooth and the oscillatory components of the spectral density. For the smooth spectral density we get (see sections 2.1, 2.2, 3.1 and appendix A for the derivation)

$$\bar{d}(k;\kappa) = \frac{Ak}{2\pi} - \frac{L}{4\pi}\left[1 - \frac{2}{\sqrt{1 + \left(\frac{\kappa}{k}\right)^2}}\right]$$
$$+ \frac{1}{4\pi k}\frac{1 + 2\left(\frac{\kappa}{k}\right)^2 - \left[1 + \left(\frac{\kappa}{k}\right)^2\right]^{3/2}}{\frac{\kappa}{k}\left[1 + \left(\frac{\kappa}{k}\right)^2\right]^{3/2}}\int_\Sigma \frac{\mathrm{d}s}{R(s)} + d_c(k;\kappa) + \ldots . \quad (7)$$

This is the generalized Weyl formula. As expected, the leading term involving the area of the billiard $A$ is independent on the boundary condition. The higher order corrections, starting from the term containing the circumference $L$, depend on $\kappa$ in a way which interpolates between the known expressions for the Dirichlet and the Neumann boundary conditions. The third term is the curvature contribution which contains an integral over the curvature $1/R(s)$, and $d_c(k;\kappa)$ denotes contributions from corners. For a 90° corner with mixed boundary conditions on one side and Dirichlet or Neumann boundary conditions on the other side it is given by

$$d_c(k;\kappa) = \mp\frac{1}{4\pi k}\frac{\frac{\kappa}{k}}{1 + \left(\frac{\kappa}{k}\right)^2} , \quad (8)$$

where the negative sign corresponds to the Dirichlet case.

The semiclassical expression for $d_{osc}(k;\kappa)$ is written as a sum of oscillatory terms whose periods correspond to the lengths of classical periodic manifolds. The semiclassical treatment distinguishes between contributions of unstable, isolated periodic orbits and contributions of continuous manifolds of neutral periodic orbits. In the standard theory, the former are given by the Gutzwiller trace formula [4, 5] and the latter were first derived by Berry and Tabor [6, 7]. To leading order, the introduction of mixed boundary conditions does not affect the amplitude of the oscillating terms. The phase of each term is changed relative to the case with Dirichlet boundary conditions, by the addition of the $\kappa$ dependent phase

$$2\sum_{i=1}^n \arctan\left(\frac{k}{\kappa(\vec{r}_i)}\cos\theta^i\right) , \quad (9)$$

where the periodic orbit bounces $n$ times off the boundary at the points $\{\vec{r}_i\}_{i=1,\cdots,n}$ and $k\cos\theta^i$ is the component of the momentum normal to the boundary at the i'th bouncing point. This expression is the same whether the periodic orbit is an isolated unstable periodic orbit (see equations (77) and (81)) or belongs to a manifold of neutral periodic orbits (see equations (37) and (47)). When $\kappa(\vec{r}) = 0$, one recovers the well known result that there is a phase difference of $n\pi$ between the contributions of periodic manifolds in the Dirichlet and Neumann cases. An important feature of (9) is that except for the limiting Dirichlet and Neumann cases the phase depends on the normal momenta at the bouncing points. We would like to emphasize that the underlying classical dynamics is indifferent to the boundary conditions. In other words, by changing the boundary condition, one can alter the quantum spectrum, without affecting the periodic orbits.

We shall explain now how the parametric dependence discussed above can be used as a diagnostic tool in the semiclassical analysis of spectra. Suppose we would like to isolate the contribution of periodic orbits which are confined to a certain segment of the boundary. We may calculate the spectrum for two $\kappa(\vec{r})$ functions which are different only in the desired segment. The difference in the oscillatory parts of the corresponding spectral densities will then depend only on trajectories which bounce off the domain of interest at least once. This is of particular importance when the boundary under consideration permits the existence of continuous families of marginally stable periodic orbits such as the "bouncing ball" orbits in the Sinai and the stadium billiards [8, 9]. Their effect is best seen when the spectral density is Fourier transformed to give a length spectrum. Here, each family contributes a peak at the lengths of the bouncing ball orbit and its repetitions. If the original spectrum is considered in the interval $[0, k_{max}]$, the intensity of the peaks at the bouncing ball lengths for a $d$-dimensional billiard scales like $k_{max}^{(d-1)/2}$ relative to the peaks appearing at the lengths of the unstable periodic orbits. Moreover, as $d$ increases the number of possible bouncing ball families proliferates and they can fill subspaces of dimensions ranging between two and $d$ in configuration space. Thus, the bouncing ball contributions may dominate the length spectrum to the extent that contributions from the unstable periodic orbits can hardly be resolved. The variation of the boundary conditions along the sections of the boundary which are visited by the unstable periodic orbits exclusively, enables us to isolate their contribution from the non-generic features which are due to the bouncing ball families. We shall show the power of this method in section 4 where we discuss in detail the spectrum of the Sinai billiard in the plane.

Another interesting application is the possibility to study the parametric spectral statistics, along the lines which were introduced in [10, 11, 12, 13] for parameter dependent Hamiltonian systems. The system that we study is different from the standard one, since the underlying classical dynamics is *independent* on the parameter $\alpha$, so that the classical periodic orbits are the same for all parameter values. The only change in the semiclassical expression is due to the phase factor (9). In particular, we shall study in section 5 the distributions of spectral "velocities" $dE_n(\alpha)/d\alpha$ and "curvatures" $d^2E_n(\alpha)/d\alpha^2$ for the Sinai billiard. Parametric statistics of this kind were previously performed for parameter dependent Hamiltonian systems and universal distributions for chaotic systems were suggested, based on random matrix models.

Much of the present work was dedicated to the derivation of the semiclassical spectral density for the mixed boundary condition. This requires some special care, as was first noted by Balian and Bloch in their work on the smooth spectral density for billiards in three dimensions with mixed boundary conditions [3]. In their analysis, Balian and Bloch explain very clearly the main difficulty which is encountered when one addresses general boundary conditions rather than the Dirichlet or Neumann conditions. Their starting point is an integral equation for the Green function, with a kernel which involves the free Green function and its normal derivative on the boundary. A Born expansion then leads to a multiple reflection series for the Green function. Balian and Bloch showed that the kernel of the integral operator is regular for Dirichlet and Neumann boundary conditions. However, for any intermediate case the kernel is singular (but still integrable). Because of this singular behavior, any perturbative expansion or semiclassical approximation, which can be used for the two extreme boundary conditions cannot be justified for any interme-diate case. Balian and Bloch observe that *"It is therefore not possible to use directly the integral equation for deriving a perturbation expansion...For instance, even for a plane, the expansion should be resummed to all orders in $\kappa$."* Balian and Bloch did not provide the corresponding treatment for the oscillatory component of the spectral density, nor did they compute the smooth spectral density for two–dimensional billiards. Moreover, their method excludes the treatment of corners. The following section will be devoted to a discussion of these points. We shall use a variety of methods to overcome the difficulty which was identified by Balian and Bloch. Sections 2 and 3 are rather technical. The reader who is not interested in the details of the derivation can skip them and, equipped with equations (7,9) he/she can go directly to the applications.

The paper is organized in the following way. In the next two sections we shall present the semiclassical theory for the spectral density. We start with integrable billiards, the circle and the rectangle, and then proceed to more general shapes which correspond to chaotic classical billiards. The numerical demonstrations and checks will be carried out for the Sinai billiard and therefore we shall give an independent derivation of the spectral density using a method which is based on the KKR technique as used by Berry [8]. In all cases, we shall discuss separately the smooth and the oscillatory parts of the spectral density. For the former, we shall use a variety of methods, extending the Stewartson and Waechter [14] calculations for the circle and the Balian and Bloch approach for smooth billiards. The resulting expression (7) for the smooth density includes the modifications due to the mixed boundary conditions for the length and curvature terms as well as a 90° corner term. To calculate the modification of the oscillating part of the spectral density we shall mostly use the scattering approach [15, 16, 17]. Section 4 will be devoted to the application of the mixed boundary conditions as a diagnostic tool. We shall develop a method for the elimination of all the structures which are due to the bouncing ball families and show how it works in practice for the Sinai billiard in two dimensions. In section 5 we shall discuss spectral statistics that depend on the sensitivity of the energy levels to a change of an external parameter. In particular, we compare numerical results to distributions that were obtained from random matrix theory, and we discuss deviations from these distributions that are due to the existence of families of bouncing ball orbits. Again, the system to be analyzed is the Sinai billiard and the boundary condition on the dispersing arc will be varied. The last section contains some discussion about the physical interpretation of the mixed boundary conditions and the behavior of the boundary phase for long unstable periodic orbits.

## 2  The Semiclassical Spectral Density – Integrable Systems

The semiclassical theory relates the oscillatory part of the spectral density to the periodic manifolds of the underlying classical dynamics. This is why it is necessary to treat separately billiards whose classical analogues are integrable or chaotic. We shall follow this route in this paper and present now the theory for two integrable billiards - the circle and the rectangle, deferring the treatment of chaotic billiards to the next section. The semiclassical expressions for the smooth components of the spectral density do not depend on the detailed classical dynamics. Because of the relative simplicity of the circle and the rectangle billiards one can use special methods to derive the smooth spectral densities in these cases. We quote them here since they provide both physical insight and independent checks for the general theory presented in the next section. Moreover, the treatment of corners will not be done by the method of Balian and Bloch and we rely on the analysis of the integrable problems for this information.

## 2.1 The Circle Billiard

We start with the circle billiard with mixed boundary conditions which is integrable if $\kappa(\vec{r})$ is constant. In this case the semiclassical approximation can be derived from the exact solutions for the Green function and the scattering matrix. Although the circle billiard is a special system, the results already show how the semiclassical trace formula will be modified in the general case.

The exact solutions of the Helmholtz equation for the circle billiard with a constant $\kappa(\vec{r})$ are given in polar coordinates by

$$\psi_{l,m}(r,\theta) = c_{l,m} \, J_l(k_{l,m} r) \, e^{il\theta} \; , \quad l \in \mathbf{Z} \; , \quad m \in \mathbf{Z}\backslash\{0\} \; , \qquad (10)$$

where $c_{l,m}$ are normalization constants and $J_l(z)$ are Bessel functions of the first kind. The wave numbers $k_{l,m}$ are determined by the boundary conditions (4). They have to satisfy

$$\kappa \, J_l(k_{l,m} R) + k_{l,m} \, J'_l(k_{l,m} R) = 0 \; . \qquad (11)$$

For $0 < \kappa < \infty$ all solutions of eq. (11) are real [18] and the positive solutions $k_{l,m}$ lie in between those for Neumann and Dirichlet boundary conditions: $k^N_{l,m} < k_{l,m} < k^D_{l,m}$ , $m > 0$. There exists an equal number of negative solutions $k_{l,-m} = -k_{l,m}$ which correspond to the same quantum state. Let us discuss briefly, what happens if $\kappa$ becomes negative. For every $l$, the first positive solution $k_{l,1}$ of eq. (11) decreases as $\kappa$ is decreased and it coincides with zero and with the first negative solution $k_{l,-1}$ if $\kappa = -l$. If $\kappa$ is decreased further, then these two solutions become a pair of complex conjugate imaginary numbers. For $l = 0$ these solutions are imaginary for any negative value of $\kappa$. Imaginary solutions of this kind exist also for other billiards and for that reason our considerations are restricted to non–negative values of $\kappa$ [1].

#### The smooth part of the spectral density

In order to obtain the smooth parts of the spectral density and of the spectral staircase for the circle billiard we apply the method of Stewartson and Waechter [14] who derived the corresponding results for Dirichlet boundary conditions. The method is based on the fact that the exact Green function of the circular billiard can be written in a closed form.

The starting point is the Green function for the heat diffusion equation which satisfies the inhomogeneous differential equation

$$(\nabla^2 - s^2) \, \tilde{G}(\vec{r}, \vec{r}\,', s^2) = \delta(\vec{r} - \vec{r}\,') \qquad (12)$$

with mixed boundary conditions on a circle of radius R:

$$(\kappa + \partial_{\hat{n}}) \, \tilde{G}(\vec{r}, \vec{r}\,', s^2) = 0 \; , \quad |\vec{r}| = R \; . \qquad (13)$$

This Green function is directly related to the Green function of the Helmholtz equation by $\tilde{G}(\vec{r},\vec{r}',s^2) = G(\vec{r},\vec{r}',k^2)|_{k=is}$ and to the spectral density by ($k>0$)

$$\begin{aligned} d(k) &= -\frac{2k}{\pi} \lim_{\nu\to 0} \text{Tr Im } \tilde{G}(\vec{r},\vec{r}',s^2-i\nu)|_{s=-ik} \\ &= -\frac{2k}{\pi} \lim_{\nu\to 0} \int d^2\vec{r}\, [\text{Im } \tilde{G}(\vec{r},\vec{r}',s^2-i\nu)|_{s=-ik}]_{\vec{r}'=\vec{r}}\,, \end{aligned} \quad (14)$$

where the integration extends over the domain of the circle. In the following an asymptotic expansion of the Green function for large $s$ will be derived and it yields an asymptotic expansion of the smooth part of the spectral density for large $k$. A detailed discussion of this point is given in [19].

The solution to equations (12) and (13) is obtained by taking the free Green function

$$\tilde{G}_0(\vec{r},\vec{r}',s^2) = -\frac{1}{2\pi} K_0(s|\vec{r}-\vec{r}'|) = -\frac{1}{2\pi} \sum_{n=-\infty}^{\infty} I_n(sr_<)K_n(sr_>)\cos[n(\theta-\theta')]\,, \quad (15)$$

which satisfies (12) and adding to it solutions of the corresponding homogeneous differential equation in order to satisfy the boundary conditions. In eq. (15) $(r,\theta)$ and $(r',\theta')$ are the polar coordinates of $\vec{r}$ and $\vec{r}'$, $r_<$ ($r_>$) is the smaller (greater) of $r$ and $r'$. $I_n(z)$ and $K_n(z)$ are modified Bessel functions [20]. The Green function $\tilde{G}(\vec{r},\vec{r}',s^2)$ can then be written as

$$\tilde{G}(\vec{r},\vec{r}',s^2) = -\frac{1}{2\pi} \sum_{n=-\infty}^{\infty} I_n(sr_<)[K_n(sr_>) + a_n I_n(sr_>)]\cos[n(\theta-\theta')]\,, \quad (16)$$

and the coefficients $a_n$ follow from the boundary condition (13) as

$$a_n = -\frac{\kappa K_n(sR) + sK'_n(sR)}{\kappa I_n(sR) + sI'_n(sR)}\,. \quad (17)$$

In order to obtain the spectral density one has to evaluate the trace of the Green function. However, since the real part of this trace is divergent one considers instead

$$\tilde{K}(s^2) = \int_0^R dr\, r \int_0^{2\pi} d\theta\, [\tilde{G}(\vec{r},\vec{r}',s^2) - \tilde{G}_0(\vec{r},\vec{r}',s^2)]_{\vec{r}'=\vec{r}}\,, \quad (18)$$

which is finite. Evaluating the two integrals results in

$$\tilde{K}(s^2) = \frac{R^2}{2} \sum_{n=-\infty}^{\infty} f(n,s)\,, \quad (19)$$

where

$$f(n,s) = \left[(1+\frac{n^2}{(sR)^2})I_n(sR)K_n(sR) - I'_n(sR)K'_n(sR) - \frac{I'_n(sR)}{sR\, I_n(sR)}\right] \frac{\kappa + s\frac{K'_n(sR)}{K_n(sR)}}{\kappa + s\frac{I'_n(sR)}{I_n(sR)}}\,. \quad (20)$$

The function $f(n,s)$ is even in the first argument for integer $n$. In order to derive the asymptotic behavior of $\tilde{K}(s^2)$ for large $s$, the sum in eq. (19) can be replaced by the integral

$$\tilde{K}(s^2) = R^2 \int_0^\infty d\nu\, f(\nu,s)\,, \quad (21)$$

since the correction terms are exponentially small for large $s$ [14]. One proceeds now in the following way. The definition (20) of $f(n, s)$ consists of two factors. The first one is the expression for $f(n, s)$ in the case of Dirichlet boundary conditions on the circle and for $R = 1$ it is identical to the definition of Stewartson and Waechter [14]. The Bessel functions in this term are replaced by their uniform approximation [20] and after substituting $\nu/(sR)$ by $\mu$, the term is expanded in powers of $1/(sR)$. The same is done with the quotients $K'_n(sR)/K_n(sR)$ and $I'_n(sR)/I_n(sR)$ in the second factor of definition (20). Taking into account terms which contribute to the first two leading terms of $K(s^2)$ yields

$$\tilde{K}(s^2) = R^2 \int_0^\infty d\mu \, \frac{1}{2sR\,(1+\mu^2)} \left[1 - \frac{\mu^2}{sR\,(1+\mu^2)^{3/2}}\right] \frac{\frac{\kappa}{s} - \sqrt{1+\mu^2} - [2sR(1+\mu^2)]^{-1}}{\frac{\kappa}{s} + \sqrt{1+\mu^2} - [2sR(1+\mu^2)]^{-1}}. \tag{22}$$

In this expression the variable $s$ appears in two different combinations, as $sR$ or $s/\kappa$. In keeping with our remarks in the introduction we expand the integrand in powers of $1/(sR)$ keeping the ratio $\kappa/s$ fixed. This leads to a resummation of $\bar{d}(k)$ to all orders in $\kappa$, which is identical to the resummation performed by the method of Balian and Bloch [3] which is reported in section 3.1. The result is

$$\tilde{K}(s^2) = R^2 \int_0^\infty d\mu \left[\frac{\frac{\kappa}{s} - \sqrt{1+\mu^2}}{2sR\,(1+\mu^2)(\frac{\kappa}{s} + \sqrt{1+\mu^2})} - \frac{-\mu^4 + \frac{\kappa^2}{s^2}\mu^2 + 1}{2(sR)^2\,(1+\mu^2)^{5/2}(\frac{\kappa}{s} + \sqrt{1+\mu^2})^2}\right]. \tag{23}$$

From this the first asymptotic terms of the smoothed level density $\bar{d}(k)$ follow as

$$\begin{aligned}\bar{d}(k) &= \frac{kR^2}{2} - \frac{2k}{\pi} \lim_{\nu \to \infty} \text{Im}\,\tilde{K}(s^2 - i\nu)\bigg|_{s=-ik} \\ &= \frac{kR^2}{2} - \frac{R}{2}\left[1 - \frac{2}{\sqrt{1+\left(\frac{\kappa}{k}\right)^2}}\right] + \frac{1}{2k}\frac{1 + 2\left(\frac{\kappa}{k}\right)^2 - \left[1+\left(\frac{\kappa}{k}\right)^2\right]^{3/2}}{\frac{\kappa}{k}\left[1+\left(\frac{\kappa}{k}\right)^2\right]^{3/2}},\end{aligned} \tag{24}$$

where the area term $kR^2/2$ is obtained from the contribution of the free Green function to the trace of $G$. The integral over $\mu$ has been performed using MAPLE. Integrating eq. (24) over $k$ yields

$$\bar{N}(k) = \frac{(kR)^2}{4} - \frac{kR}{2}\left[1 - 2\left(\sqrt{1+\left(\frac{\kappa}{k}\right)^2} - \frac{\kappa}{k}\right)\right] + \frac{1}{6}\left[1 - 3\frac{\sqrt{1+\left(\frac{\kappa}{k}\right)^2} - 1}{\frac{\kappa}{k}\sqrt{1+\left(\frac{\kappa}{k}\right)^2}}\right]. \tag{25}$$

In principle, the integration of $\bar{d}(k)$ over $k$ could lead to a $\kappa$-dependent constant in the asymptotic expression of $\bar{N}(k)$. One way to determine the constant term of the asymptotic $\bar{N}(k)$ is to expand $-\tilde{K}(s^2)$ in all orders of $1/s$ (now taking into account also the $s/\kappa$-terms). Then the coefficient of the $1/s^2$-term is identical to the constant term in the asymptotic series for $\bar{N}(k)$ [21], if expanded in all powers of $1/k$. Doing this, one obtains in both cases $1/6 - \kappa R$, and thus the constant term in eq. (25) is correct. Eq. (25) is identical to the formula for general shapes that will be derived in section 3, if the three terms are expressed by the area $A = \pi R^2$, the perimeter $L = 2\pi R$ and the integrated curvature $\int \mathrm{d}s \, / \, R(s) = 2\pi$, respectively.

In appendix A the corresponding result for a semi-circle billiard is derived with mixed boundary conditions on the circle and Dirichlet or Neumann boundary conditions on the diameter. In this way the contribution of a 90° corner is obtained that will be used in applications in section 3.3.

In figure 1 we show a comparison of the three contributions to the asymptotic expression (25) for the mean spectral staircase $\bar{N}(k)$ with numerical data for the circle billiard with $\kappa = 150$. All 22 540 eigenvalues below wave number $k = 300$ have been included. In the upper figure the area term is compared to the numerical result for $N(k)$ and on this scale one cannot see any difference. The intermediate figure shows the length term in comparison with the difference between the spectral staircase and the area term. The numerical data have been smoothed over the range of approximately 100 levels. Both curves are in good agreement over the whole range of $k$. In order to demonstrate the effect of the resummation in $\kappa$, we also plotted the length term for Neumann boundary conditions (dashed), since this is the next-to-leading asymptotic term for all values of $\kappa$, if no resummation is performed. The dashed and the full line become parallel for large values of $k$, but one can see that the resummation is necessary in order to have a good description of the spectrum also for lower values of $k$. In the lower figure the curvature term is compared to the spectral staircase from which the area and length terms have been subtracted. In this case the numerical curve has to be smoothed strongly (over the range of approximately 1000 levels) since it has large fluctuations about its mean (note the different scales in the three figures). Nevertheless, one gets good agreement with the theoretical curve. The deviations near $k = 300$ and $k = 0$ (also in the intermediate figure) are due to edge effects of the smoothing procedure.

## The oscillatory part of the spectral density

We derive the oscillatory part of the spectral density by applying a scattering formalism and making use of the fact that there is a direct relation between the inside and outside problem of a billiard system.

A solution of the scattering problem for a compact billiard system can outside a circle that contains the billiard completely be written as

$$\psi(r, \theta) = H_l^-(kr) \, i^l e^{il\theta} + \sum_{l'=-\infty}^{\infty} S_{l,l'}(k) H_{l'}^+(kr) \, i^{l'} e^{il'\theta} \, , \tag{26}$$

where $S_{l,l'}(k)$ are the elements of the on-shell scattering matrix in the angular momentum representation, and $H_l^\pm(z) = J_l(z) \pm iY_l(z)$ are Hankel functions. For the circle (with radius $R$), the requirement that $\psi(r, \theta)$ satisfies the boundary conditions (4) with constant $\kappa$ gives the scattering matrix explicitly. Due to the rotational invariance of the system, the scattering matrix is diagonal, and its diagonal elements are given by

$$S_l(k) = e^{-i\varphi_l(k)} = -\frac{\kappa \, H_l^-(kR) + k \, H_l^{-'}(kR)}{\kappa \, H_l^+(kR) + k \, H_l^{+'}(kR)} \, , \tag{27}$$

where $\varphi_l(k)$ are the scattering phases. From this equation, and from the eigenvalue equation for the inside problem (11), it follows that

$$S_l(k) = 1 \iff k \in \{k_{l,n} \mid n = 1, 2, \ldots\} \tag{28}$$

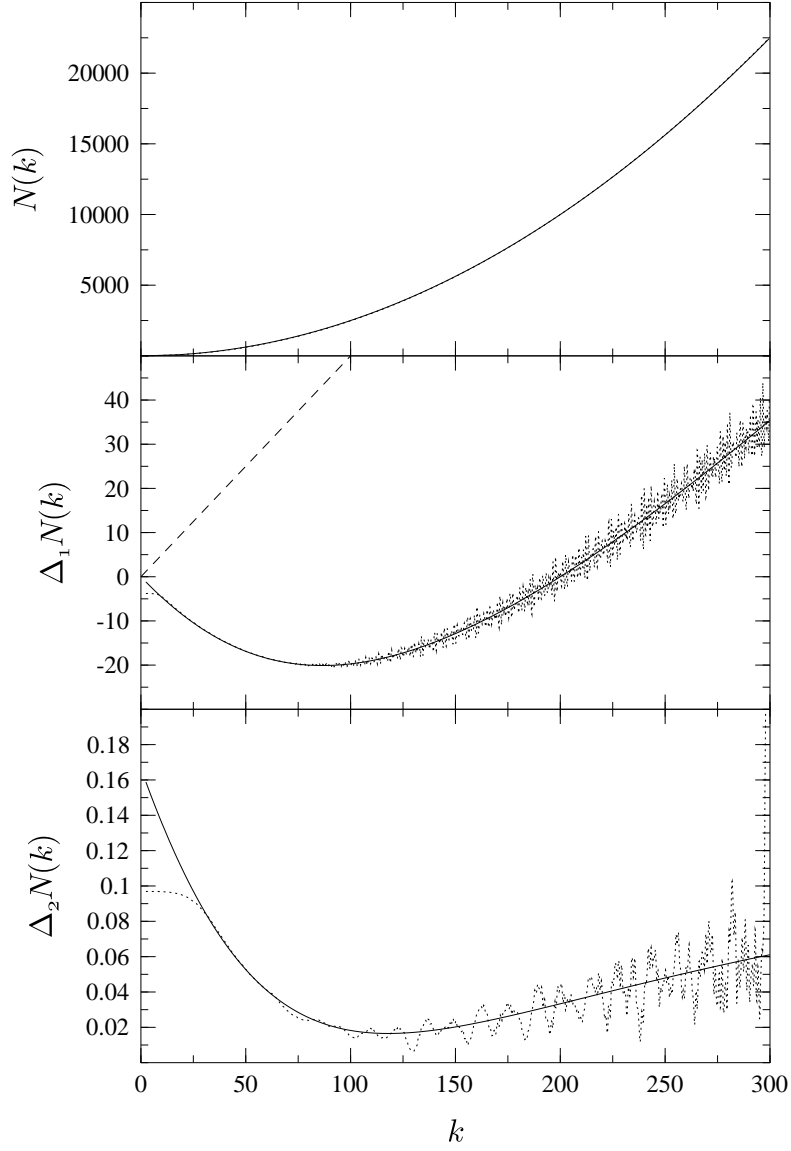

Figure 1: The area, length and curvature term of the mean spectral staircase $\bar{N}(k)$ (full lines) for the circular billiard with $\kappa = 150$ in comparison with the spectral staircase $N(k)$ from which the respectively stronger asymptotic terms have been subtracted (dotted lines). In the second and third figure the dotted lines have been obtained by smoothing the original curves. The dashed line shows the length term for Neumann boundary conditions.

for all $l \in \mathbf{Z}$. This is a direct relation between the inside problem and the scattering problem, and is a particular example of the inside-outside duality for billiard systems [22], which has been proved in a general form recently [17].

The semiclassical approximation for the $S$-matrix is obtained by using the Debye approximation for the Hankel functions (or equivalently by a WKB approximation). This approximation is valid for $l < kR$, with the further requirement that $|l - kR|$ is larger than approximately $(kR)^{1/3}$. These conditions have consequences on the accuracy of the semiclassical trace formula which is derived from it. The first condition is related to the fact that the semiclassical approximation will be given in terms of the classical orbits which satisfy $l < kR$. The consequence of the second condition is that the contribution of orbits to the trace formula becomes less accurate, as the angle of incidence approaches $\pi/2$ (whispering gallery orbits). An improvement of the approximation is possible by using a uniform approximation for the Hankel functions [23].

Inserting the Debye approximation for the Hankel functions and their derivatives into eq. (27) we obtain

$$S_l(k) = e^{-i\varphi_l(k)} \approx -\frac{(\kappa - ip_l)\, e^{-i(\sigma_l - \pi/4)}}{(\kappa + ip_l)\, e^{+i(\sigma_l - \pi/4)}} = \exp\left[-i\left(2\sigma_l - \frac{3\pi}{2} + 2\arctan\left(\frac{p_l}{\kappa}\right)\right)\right], \quad (29)$$

where

$$\sigma_l = k\sqrt{R^2 - l^2/k^2} - |l|\arccos\frac{|l|}{kR} \quad \text{and} \quad p_l = \sqrt{k^2 - l^2/R^2}\,. \quad (30)$$

Eq. (29) determines the semiclassical scattering phases $\varphi_l(k)$. We now make use of the inside-outside duality in order to obtain from these phases a semiclassical approximation for the density of states of the inside problem. From relation (28) it follows that $(k > 0)$

$$d(k) = \sum_{l=-\infty}^{\infty} \sum_{m=1}^{\infty} \delta(k - k_{l,m}) = \sum_{l=-\infty}^{\infty} \delta_p(\varphi_l(k)) \left|\frac{\partial\varphi_l(k)}{\partial k}\right|, \quad (31)$$

where $\delta_p$ is a periodic delta-function with period $2\pi$. The wave numbers for the circle satisfy $l < kR$, and since in this region the derivatives $\partial\varphi_l(k)/\partial k$ are positive, the absolute value can be omitted from eq.(31).

Now the delta-function $\delta_p$ is expressed as a sum over exponential functions. Neglecting the constant term in this sum, which contributes to the smooth part of the level density, we obtain

$$d_{osc}(k) = \frac{1}{2\pi} \sum_{n \neq 0} \sum_{l=-\infty}^{\infty} \frac{\partial\varphi_l(k)}{\partial k}\, e^{-in\varphi_l(k)} = -\frac{1}{\pi}\,\mathrm{Im}\sum_{n=1}^{\infty} \frac{1}{n}\,\mathrm{Tr}\frac{\partial}{\partial k}S^n(k)\,. \quad (32)$$

This is equal to the general result for $d_{osc}(k)$ in the scattering approach [15].

The trace of the derivative of the $n$-th power of the $S$-matrix is semiclassically calculated by applying the Poisson resummation formula

$$\mathrm{Tr}\frac{\partial}{\partial k}S^n(k) = -i\,n\sum_{m=-\infty}^{\infty}\int_{-\infty}^{\infty}\mathrm{d}l\,\frac{\partial\varphi_l(k)}{\partial k}e^{-in\varphi_l(k) - 2\pi iml}\,, \quad (33)$$

and evaluating the integral over $l$ by a stationary phase approximation.

The derivative with respect to $l$ of the arctan–term in eq. (29) is of order $1/kR$ smaller than $d\sigma_l/dl$ when $|l - kR| \gg (kR)^{1/3}$. It remains smaller in comparison with $d\sigma_l/dl$ when $|l - kR|$ approaches $(kR)^{1/3}$, and can therefore be neglected in the region of validity of the Debye approximation. The stationary phase condition for $l$ is then given by

$$l_{n,m} = kR \operatorname{sign}(m) \cos\left(\frac{\pi m}{n}\right) . \tag{34}$$

This condition leads to a restriction of the sum over $m$ to values for which $|m| \leq n/2$. For $|m| = n/2$ the stationary point is at $l = 0$ where $\partial \varphi_l / \partial l$ is discontinuous, and the integration therefore is evaluated only on one side of the stationary point, which leads to an additional factor of $1/2$.

The numbers $n$ and $m$ can be given a physical interpretation. The stationary points of eq. (34) correspond to the classical periodic orbits of the circle billiard, which are characterized by the number of reflections on the boundary $n$ and the winding number $m$. The saddle point $l_{n,m}$ is the angular momentum of the orbit.

Performing the stationary phase integration and inserting the result into eq. (32) finally gives

$$d_{osc}(k) = \sum_{n=2}^{\infty} \sum_{m=1}^{[n/2]} g_m \sqrt{\frac{4kR^3}{\pi n} \sin^3\left(\frac{\pi m}{n}\right)} \cos\left[2nkR \sin\left(\frac{\pi m}{n}\right)\right.$$
$$\left. -n\frac{3\pi}{2} + \frac{\pi}{4} + 2n \arctan\left(\frac{k}{\kappa} \sin\left(\frac{\pi m}{n}\right)\right)\right] , \tag{35}$$

where

$$g_m = \begin{cases} 1 & m = n/2 \\ 2 & m \neq n/2 \end{cases} . \tag{36}$$

This formula expresses the oscillatory part of the spectral density in terms of a summation over all families of periodic orbits, which are denoted by the two integers $n$ and $m$. Its form agrees with the general semiclassical formula for integrable systems of Berry and Tabor [6, 7], but it has additional phases due to the boundary conditions. The first term in the cosine is the action along the periodic orbit $kL_{n,m}$, where $L_{n,m} = 2nR \sin(\pi m/n)$ is the orbits length. The phase $-3\pi n/2$ is due to $n$ reflections on the boundary and $n$ conjugate points along the orbit. In comparison to the result for Dirichlet boundary conditions there is an additional phase that consists of a contribution of

$$2 \arctan\left(\frac{k}{\kappa} \sin\left(\frac{\pi m}{n}\right)\right) \tag{37}$$

for every reflection on the boundary, where $\sin(\pi m/n)$ is identical to the cosine of the angle of incidence. This is the phase that is given in eq. (9) in the introduction.

## 2.2 The Rectangle Billiard

Quantization of the rectangle billiard with mixed boundary conditions has a twofold purpose. First, it serves as a check for the generality of the results that were obtained for the circle and the semi-circle billiards with mixed boundary conditions. Second, as the derivation unfolds, we arrive at a useful expression that can be used to calculate the semi-classical spectral density for a class of higher–dimensional billiards with mixed boundary conditions.

For simplicity, we quantize a $L_x \times L_y$ rectangle, with Dirichlet boundary conditions on the edges of length $L_x$ and mixed boundary conditions on the edges of length $L_y$ (see figure 2). Due to separability, we readily get the quantization conditions

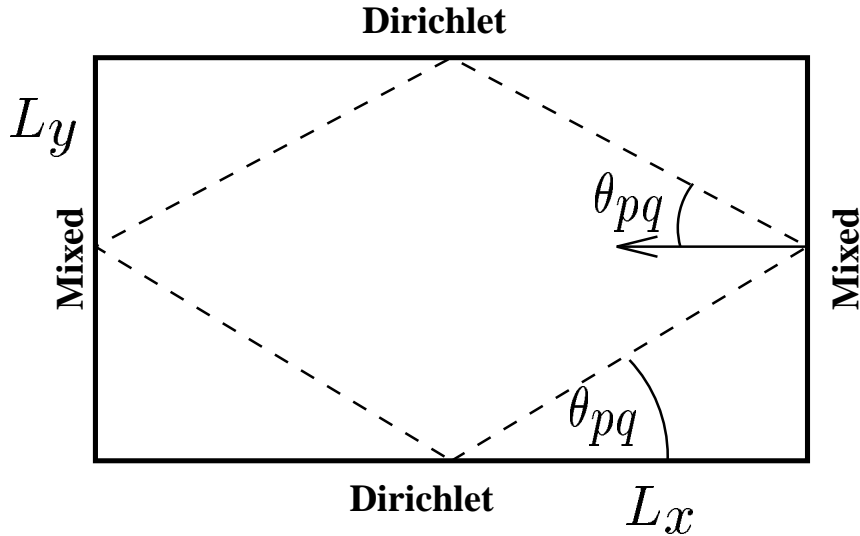

Figure 2: General definitions for the rectangle billiard.

$$k_{x,n} L_x + 2\arctan\left(\frac{k_{x,n}}{\kappa}\right) = n\pi, \quad n = \pm 1, \pm 2, \ldots \tag{38}$$

and

$$k_{y,m} L_y = m\pi, \quad m = \pm 1, \pm 2, \ldots, \tag{39}$$

where $k^2 = k_x^2 + k_y^2$ and $(\pm n, \pm m)$ represent the same quantum state. The left hand side of (38) is monotonically increasing in $k_{x,n}$ from $-\infty$ to $+\infty$ and thus guaranteeing a unique solution for every $n$ (which justifies the notation). Similar considerations also apply to eq. (39).

The spectral density for the rectangle can be written as

$$d^{(2D)}(E) = \sum_{n,m=1}^{\infty} \delta(E - E_{nm})$$
$$= \frac{1}{4}\left[\sum_{n,m=-\infty}^{\infty} \delta(E - E_{nm}) - \sum_{n=-\infty}^{\infty} \delta(E - E_{n0}) - \sum_{m=-\infty}^{\infty} \delta(E - E_{0m}) + \delta(E - E_{00})\right], \tag{40}$$

where $E = k^2$, $E_{nm} = k_{nm}^2 = k_{x,n}^2 + k_{y,m}^2$ and we took advantage of the antisymmetry relations $k_{x,-n} = -k_{x,n}$, $k_{y,-m} = -k_{y,m}$ which are easily derived from the quantization conditions (38, 39). In particular, $E_{00} = 0$. Apply Poisson summation to the first term in (40), using the natural continuations of (38, 39) to real $m, n$:

$$\sum_{n,m=-\infty}^{\infty} \delta(E - E_{nm}) = \sum_{n,m=-\infty}^{\infty} \delta\left(E - k_{x,n}^2 - k_{y,m}^2\right)$$

$$= \sum_{p,q=-\infty}^{\infty} \int_{-\infty}^{\infty} \mathrm{d}n\, \mathrm{d}m\, \delta\left[E - k_x^2(n) - k_y^2(m)\right] e^{2\pi i (pn+qm)}$$

$$= \sum_{p,q=-\infty}^{\infty} \int_{-\infty}^{\infty} \mathrm{d}k_x\, \mathrm{d}k_y\, \frac{\mathrm{d}n}{\mathrm{d}k_x}\frac{\mathrm{d}m}{\mathrm{d}k_y} \delta\left(E - k_x^2 - k_y^2\right) e^{2\pi i [pn(k_x)+qm(k_y)]}$$

$$= \frac{1}{2} \sum_{p,q=-\infty}^{\infty} \int_0^{2\pi} \mathrm{d}\theta\, n'(k\cos\theta)\, m'(k\sin\theta)\, e^{2\pi i [pn(k\cos\theta)+qm(k\sin\theta)]} \,. \quad (41)$$

The first change of integration variables $(m, n) \to (k_x, k_y)$ is allowed due to the monotonicity of $n(k_x)$, $m(k_y)$ mentioned above. The second change of variables is just transforming to polar coordinates $(k, \theta)$, such that the $k$ integration can be explicitly performed, eliminating the $\delta$ term. The primes in the above expression denote differentiation with respect to the argument. Similar considerations applied to the other terms of (40) finally lead to

$$d^{(2D)}(E) = \frac{1}{8} \sum_{p,q=-\infty}^{\infty} \int_0^{2\pi} \mathrm{d}\theta\, \tilde{d}_{x,p}^{(1D)}(k\cos\theta)\tilde{d}_{y,q}^{(1D)}(k\sin\theta) e^{2\pi i [pn(k\cos\theta)+qm(k\sin\theta)]} \,, \quad (42)$$

where

$$\tilde{d}_{x,p}^{(1D)}(k) = n'(k) - \delta_{p,0}\delta(k)\,, \quad \tilde{d}_{y,q}^{(1D)}(k) = m'(k) - \delta_{q,0}\delta(k)\,. \quad (43)$$

For $p = q = 0$ eq. (42) gives the smooth two–dimensional spectral density

$$\bar{d}^{(2D)}(E) = \frac{1}{8} \int_0^{2\pi} \mathrm{d}\theta\, \bar{d}_x^{(1D)}(k\cos\theta)\bar{d}_y^{(1D)}(k\sin\theta)\,, \quad (44)$$

where $\bar{d}_x^{(1D)}(k) = n'(k) - \delta(k)$ and $\bar{d}_y^{(1D)}(k) = m'(k) - \delta(k)$ are the one–dimensional smooth spectral densities (the $\delta$–functions are a convenient way to represent the constant term of $\bar{N}^{(1D)}(k)$). The "combination formulæ" (42, 44) thus express the two–dimensional spectral density using the one–dimensional components in a simple way and rely on the energy decomposition relation $E = k_x^2 + k_y^2$ and the positive–negative antisymmetry of $k_{x,\pm n}, k_{y,\pm m}$. Consequently, it can be extended to calculations of the spectral density for higher–dimensional billiards, provided that the energy can be decomposed in the above way. In particular, it is useful for three–dimensional "cylindrical" billiards [21], i.e., two–dimensional billiards elongated along the perpendicular axis and closed from above and below. For example, the cubic billiard with a variety of mixed boundary conditions can be easily calculated this way, including the smooth contributions of right angle edges and corners, which are otherwise difficult to estimate using more general methods. If one assumes that the Weyl series has the general form with volume term, surface term et cetera, also in the case of mixed boundary conditions, then these results can be applied also to billiards other than the simple ones mentioned above.

To derive semiclassical expression for $d^{(2D)}(E)$, we use the saddle point approximation to evaluate the oscillatory integrals $((p, q) \neq (0, 0))$ in (41). The phase appearing in these integrals can be divided into two parts:

$$\begin{aligned}
\varphi^{pq}(\theta) &= 2\pi\left[pn(k\cos\theta)+qm(k\sin\theta)\right] \\
&= 2\left[pkL_x\cos\theta + 2p\arctan\left(\frac{k\cos\theta}{\kappa}\right)+qkL_y\sin\theta\right] \\
&= \varphi_D^{pq}(\theta)+4p\arctan\left(\frac{k\cos\theta}{\kappa}\right) = \varphi_D^{pq}(\theta)+\phi^p(\theta) ,
\end{aligned} \quad (45)$$

where $\varphi_D^{pq}(\theta)$ is the phase that appears in the pure Dirichlet case and is rapidly oscillating in $\theta$ for large $k$. The phase $\phi^p(\theta)$, which is genuine to the mixed case, is both bounded and slowly oscillating and thus can be appended to the slowly varying prefactors as discussed in subsection 2.1 (see eq. (34)). Thus, the saddle points are the same as for the pure Dirichlet case:

$$\tan\theta_{pq}^{1,2} = \frac{pL_y}{qL_x} . \quad (46)$$

The angles $\theta_{pq}^{1,2}$ correspond to the angles between the $x$–axis and classical periodic orbits (tori) which trace the $x$ dimension $p$ times and the $y$ dimension $q$ times (see figure 2). Putting together the above results, we finally get the semiclassical approximation to $d^{(2D)}(E)$:

$$\begin{aligned}
d^{(2D)}(E) &= \bar{d}^{(2D)}(E)+d_{osc}^{(2D)}(E) \\
&= \frac{L_xL_y}{4\pi}-\frac{L_x}{4\pi k}-\frac{L_y}{4\pi k}+\frac{L_y}{2\pi k\sqrt{1+\left(\frac{\kappa}{k}\right)^2}}-\frac{\left(\frac{\kappa}{k}\right)}{2\pi k^2\left(1+\left(\frac{\kappa}{k}\right)^2\right)}+\frac{1}{4}\delta(E) \\
&+ \frac{L_xL_y}{4\pi}\sum_{p,q\neq(0,0)}\sqrt{\frac{2}{\pi kL_{pq}}}\cos\left[kL_{pq}+4|p|\arctan\left(\frac{k}{\kappa}\cos\theta_{pq}\right)-\frac{\pi}{4}\right] \\
&- \frac{L_x}{2\pi k}\sum_{p=1}^{\infty}\cos\left[2kpL_x+4p\arctan\left(\frac{k}{\kappa}\right)\right] - \frac{L_y}{2\pi k}\sum_{q=1}^{\infty}\cos(2kqL_y) ,
\end{aligned} \quad (47)$$

where $L_{pq} = 2\sqrt{(pL_x)^2+(qL_y)^2}$ is the length of the $(p,q)$ periodic orbit and $\cos\theta_{pq} = |\cos\theta_{pq}^{1,2}| = 2|p|L_x/L_{pq}$. In the smooth part we kept terms of all orders, since we evaluate the integrals exactly, but in the oscillatory part we kept only leading order terms to be consistent with the saddle point approximation. The double sum relates to the periodic tori, while the two single sums relate to "boundary orbits" [9] that bounce along the edges of the rectangle. Examining the above expression, we conclude that the effect of the mixed boundary conditions on the oscillatory part is simply to modify the contribution of each periodic orbit by a phase, which is $2\arctan[(k/\kappa)\cos\theta]$ for each bounce from a wall with mixed boundary conditions, where $\theta$ is the angle of incidence with respect to the inside–pointing normal. The same rule also applies to the boundary orbits, for which $\theta$ is either 0 or $\pi/2$. This result, as well as the modifications to the smooth spectral density (length term and right angle Dirichlet–mixed corners) are equivalent to and consistent with the results that were obtained above for the circle (see (24,37)).

# 3 General billiard systems

We now consider general smooth two–dimensional billiards. A very convenient method for the treatment of billiard systems are boundary integral methods since they reduce the two–dimensional billiard problem to an effectively one–dimensional problem along the boundary of the billiard. However, as has been discussed in the introduction, in the case of mixed boundary conditions one encounters the difficulty that the integral kernels have a singularity, which makes a direct perturbation expansion impossible. One possibility to solve this problem, is to apply a transformation to the integral equation, which leads to an non-singular kernel. This corresponds to a resummation of the equation in all orders of $\kappa$ and we will use this method in order to derive the smooth part of the level density [3]. For the oscillatory part of the level density, however, we choose two different approaches that avoid the problem of singular integral kernels. We first apply methods of scattering theory in order to obtain the semiclassical approximation for convex chaotic billiard systems. For the calculation of the scattering matrix the Kirchhoff approximation (presented in [15]) is extended to mixed boundary conditions. We further derive the semiclassical approximation for a concave billiard, the Sinai billiard, by the KKR-method.

## 3.1 The smooth part of the spectral density

In reference [3] Balian and Bloch derive the asymptotic form of $\bar{N}(E)$ for an arbitrary smooth three–dimensional billiard with mixed boundary conditions. We use their method in order to obtain the corresponding two–dimensional result. Our notation differs from that of Balian and Bloch in that the Green function and the normal derivative have a different sign.

Consider a two–dimensional domain $D$ with area $A$ and a smooth boundary $\Sigma$ of length $L$. The Green function of the Helmholtz equation for this domain is determined by the differential equation

$$\left(\nabla^2 + E\right) G(\vec{r}, \vec{r}\,', E) = \delta(\vec{r} - \vec{r}\,') \tag{48}$$

with the boundary condition

$$(\kappa + \partial_{\hat{n}}) G(\vec{r}, \vec{r}\,', E) = 0 \,, \quad \vec{r} \in \Sigma \,. \tag{49}$$

It is related to the smoothed spectral density $d_\nu(E)$, which has Lorentzian peaks of width $\nu$ at the energy values $E_n$, by

$$d_\nu(E) = -\frac{1}{\pi} \int_A \mathrm{d}^2 \vec{r} \, \left[\mathrm{Im}\, G(\vec{r}, \vec{r}\,', E + i\nu)\right]_{\vec{r}\,' = \vec{r}} \,. \tag{50}$$

For the derivation of the large $k$-behavior of the Green function the differential equation (48) and boundary condition (49) are replaced by an integral equation. This is done by representing the Green function $G(\vec{r}, \vec{r}\,', E)$ as the sum of the free Green function and a *single layer potential*:

$$G(\vec{r}, \vec{r}\,') = G_0(\vec{r}, \vec{r}\,') + \int_\Sigma \mathrm{d}s_\alpha \, G_0(\vec{r}, \vec{r}_\alpha) \mu(\vec{r}_\alpha, \vec{r}\,') \tag{51}$$

with a density $\mu(\vec{r}_\beta, \vec{r}\,')$ which is determined by

$$\mu(\vec{r}_\beta, \vec{r}\,') = 2(\kappa + \partial_{\hat{n}_\beta}) G_0(\vec{r}_\beta, \vec{r}\,') + 2 \int_\Sigma \mathrm{d}s_\alpha \, [\kappa + \partial_{\hat{n}_\beta}] G_0(\vec{r}_\beta, \vec{r}_\alpha) \mu(\vec{r}_\alpha, \vec{r}\,') \,. \tag{52}$$

For brevity of notation, the energy dependence of the Green functions and the density $\mu$ will be omitted, and Greek indices will denote coordinates on the boundary.

For Neumann boundary conditions ($\kappa = 0$) the integral equation (52) can be solved by iteratively replacing the potential $\mu$ in the integral term of eq. (52) by the whole expression for $\mu$. For $\kappa \neq 0$ this cannot be done since the Green function $G_0(\vec{r}_\beta, \vec{r}_\alpha)$, in contrast to its normal derivative, has a (logarithmic) singularity at $\vec{r}_\alpha = \vec{r}_\beta$. For that reason, the integral equation has to be transformed into a different integral equation, for which the kernel is uniformly bounded. This is done by introducing an auxiliary Green function $\Gamma(\vec{r}_\alpha, \vec{r}_\beta)$ which is defined on the boundary $\Sigma$ by

$$\Gamma(\vec{r}_\alpha, \vec{r}_\beta) - 2\kappa \int_\Sigma \mathrm{d}s_\gamma \, G_0(\vec{r}_\alpha, \vec{r}_\gamma) \, \Gamma(\vec{r}_\gamma, \vec{r}_\beta) = \delta(\vec{r}_\alpha - \vec{r}_\beta) \,. \tag{53}$$

Multiplying eq. (52) by $\Gamma(\vec{r}_\alpha, \vec{r}_\beta)$ and integrating over $\vec{r}_\beta$ one obtains

$$\begin{aligned} \mu(\vec{r}_\alpha, \vec{r}\,') &= 2 \int \mathrm{d}s_\beta \, \Gamma(\vec{r}_\alpha, \vec{r}_\beta) \, [\kappa + \partial_{\hat{n}_\beta}] G_0(\vec{r}_\beta, \vec{r}\,') \\ &\quad + 2 \int \mathrm{d}s_\beta \, \mathrm{d}s_\gamma \, \Gamma(\vec{r}_\alpha, \vec{r}_\beta) \, \partial_{\hat{n}_\beta} G_0(\vec{r}_\beta, \vec{r}_\gamma) \, \mu(\vec{r}_\gamma, \vec{r}\,') \,. \end{aligned} \tag{54}$$

This equation is now solved by a perturbation expansion and the result is inserted into eq. (51). One obtains

$$G(\vec{r}, \vec{r}\,') = G_0(\vec{r}, \vec{r}\,') + 2 \int_\Sigma \mathrm{d}s_\alpha \, \mathrm{d}s_\beta \, G_0(\vec{r}, \vec{r}_\alpha) \, \Gamma(\vec{r}_\alpha, \vec{r}_\beta) \, (\kappa + \partial_{\hat{n}_\beta}) G_0(\vec{r}_\beta, \vec{r}\,') \tag{55}$$

$$+ 4 \int_\Sigma \mathrm{d}s_\alpha \mathrm{d}s_\beta \mathrm{d}s_\gamma \mathrm{d}s_\delta \, G_0(\vec{r}, \vec{r}_\alpha) \, \Gamma(\vec{r}_\alpha, \vec{r}_\beta) \, \partial_{\hat{n}_\beta} G_0(\vec{r}_\beta, \vec{r}_\gamma) \, \Gamma(\vec{r}_\gamma, \vec{r}_\delta) \, (\kappa + \partial_{\hat{n}_\delta}) G_0(\vec{r}_\delta, \vec{r}\,') + \ldots$$

Inserting this into eq. (50) yields the following expression for the level density:

$$\begin{aligned} d_\nu(E) &= \frac{A}{4\pi} - \frac{1}{\pi} \mathrm{Im} \left[ 2 \int_\Sigma \mathrm{d}s_\alpha \, \mathrm{d}s_\beta \, \Gamma(\vec{r}_\alpha, \vec{r}_\beta) \, (\kappa + \partial_{\hat{n}_\beta}) F(\vec{r}_\beta, \vec{r}_\alpha) \right. \\ &\quad \left. + 4 \int_\Sigma \mathrm{d}s_\alpha \, \mathrm{d}s_\beta \, \mathrm{d}s_\gamma \, \mathrm{d}s_\delta \, \Gamma(\vec{r}_\alpha, \vec{r}_\beta) \, \partial_{\hat{n}_\beta} G_0(\vec{r}_\beta, \vec{r}_\gamma) \, \Gamma(\vec{r}_\gamma, \vec{r}_\delta) \, (\kappa + \partial_{\hat{n}_\delta}) F(\vec{r}_\delta, \vec{r}_\alpha) + \ldots \right] \,, \end{aligned} \tag{56}$$

where the function $F(\vec{r}_\alpha, \vec{r}_\beta)$ is defined by

$$F(\vec{r}_\alpha, \vec{r}_\beta) = \int_A \mathrm{d}^2\vec{r} \, G_0(\vec{r}_\alpha, \vec{r}) \, G_0(\vec{r}, \vec{r}_\beta) \,, \tag{57}$$

and the energy of the Green functions is given by $E + i\nu$. In contrast to the three-dimensional calculation, the introduction of a convergence factor is not necessary here since the integrals are convergent.

**The perimeter term**

The evaluation of the integrals in eq. (56) is based on the fact that the functions $G_0$ and $\Gamma$ are short-range functions when the energy $E$ is large. In a first approximation it is assumed that the range is smaller than the smallest radius of curvature of the boundary $\Sigma$ and the boundary is locally replaced by its tangent. In this approximation only the first integral term in eq. (56) contributes, since $\partial_{\hat{n}_\beta} G_0(\vec{r}_\beta, \vec{r}_\gamma)$ vanishes if the curvature of the boundary is zero. One proceeds in the following way: the coordinates $\vec{r}_\alpha$ and $\vec{r}_\beta$ are replaced by

$$\vec{r}_\xi = \vec{r}_\alpha - \vec{r}_\beta, \quad \vec{r}_\omega = \frac{1}{2}(\vec{r}_\alpha + \vec{r}_\beta) \tag{58}$$

and the integral over $\vec{r}_\xi$ is evaluated along the tangent of the boundary at the point $\vec{r}_\omega$. Along this tangential line the functions $G_0$ and $\Gamma$ are translational invariant and for that reason it is convenient to introduce the Fourier transform along the tangent:

$$\hat{f}(p) = \int ds\, e^{-ips} f(s), \quad f(s) = \frac{1}{2\pi} \int dp\, e^{ips} \hat{f}(p). \tag{59}$$

In terms of the Fourier transforms the contribution to the level density can be written as

$$d_\nu(E) = \frac{A}{4\pi} - \frac{2}{\pi^2} \text{Im} \left[ \int_\Sigma ds_\omega \int_0^\infty dp\, \hat{\Gamma}(p) \left( \kappa \hat{F}(p) + \widehat{\partial_{\hat{n}} F}(p) \right) \right] + \ldots \tag{60}$$

The calculation of the Fourier transforms of the functions $G_0$, $\Gamma$ and $F$ is done exactly as in the three–dimensional case of Balian and Bloch. For that reason, we give here only the results:

$$\hat{G}_0(p) = -\frac{1}{2a(p)}, \quad \hat{\Gamma}(p) = \frac{a(p)}{\kappa + a(p)}, \tag{61}$$

where $a(p) = \sqrt{p^2 - k_\nu^2}$, $k_\nu = \sqrt{E + i\nu}$ and

$$\hat{G}_0(p, z) = -\int_{-\infty}^\infty \frac{dq}{2\pi} \frac{e^{iqz}}{a^2(p) + q^2}, \quad \hat{F}(p) = \frac{1}{8a^3(p)}, \quad \widehat{\partial_{\hat{n}} F}(p) = -\frac{1}{8a^2(p)}. \tag{62}$$

Those of the functions in the above equations which depend only on one parameter are the Fourier transforms of functions with arguments in the plane, which are translationally invariant along the plane. $\hat{G}_0(p, z)$ is the Fourier transform of $G_0(\vec{r}_\alpha, \vec{r})$ and $z$ is the distance of $\vec{r}$ from the plane. The expressions in equations (61) and (62) have exactly the same form as in the three–dimensional case, the difference being that $p$ is now a scalar instead of a two–dimensional vector. The results (61) and (62) are inserted into eq. (60) and the limit $\nu \to 0$ is performed, yielding the first asymptotic terms for the mean level density. Multiplying the result by $2k$ gives the result for the mean level density in terms of the wave number $k$:

$$\begin{aligned}\bar{d}(k) &= \frac{A}{2\pi} k - \frac{Lk}{2\pi^2} \left[ \text{Im} \int_{-ik_\nu}^\infty da \frac{1}{a\sqrt{a^2 + k_\nu^2}} \frac{\kappa - a}{\kappa + a} \right]_{\nu \to 0} + \ldots \\ &= \frac{A}{2\pi} k - \frac{L}{4\pi} \left[ 1 - \frac{2}{\sqrt{1 + \left(\frac{\kappa}{k}\right)^2}} \right] + \ldots. \end{aligned} \tag{63}$$

**The curvature term**

The next term in the asymptotic expansion of $\bar{d}(k)$ is obtained by including those corrections to the tangent approximation which are linear in $1/R_\omega$, where $R_\omega$ is the radius of curvature at $\vec{r}_\omega$. Such corrections exist for $F(\vec{r}_\alpha, \vec{r}_\beta)$, $\partial_{\hat{n}_\alpha} F(\vec{r}_\alpha, \vec{r}_\beta)$ and $\partial_{\hat{n}_\alpha} G_0(\vec{r}_\alpha, \vec{r}_\beta)$, whereas for $G_0(\vec{r}_\alpha, \vec{r}_\beta)$ and $\Gamma(\vec{r}_\alpha, \vec{r}_\beta)$ the first corrections are quadratic in $1/R_\omega$ and can be neglected. The integrals in eq. (56) are expressed in terms of the Fourier transforms of the Green functions and their corrections and the next asymptotic term for the energy level density follows as

$$-\frac{2}{\pi^2}\mathrm{Im}\left[\int_\Sigma \mathrm{d}s_\omega \int_0^\infty \mathrm{d}p\, \hat{\Gamma}(p)\left[\kappa\, \delta\hat{F}(p) + \delta\widehat{\partial_{\hat{n}} F}(p)\right]\right]$$
$$-\frac{4}{\pi^2}\mathrm{Im}\left[\int_\Sigma \mathrm{d}s_\omega \int_0^\infty \mathrm{d}p\, \hat{\Gamma}(p)\, \delta\widehat{\partial_{\hat{n}} G_0}(p)\, \hat{\Gamma}(p)\left[\kappa\, \hat{F}(p) + \widehat{\partial_{\hat{n}} F}(p)\right]\right]. \qquad (64)$$

The Fourier transforms of the Green functions and their corrections can again be derived exactly as in the three-dimensional case [3] and give the result

$$\delta\widehat{\partial_{\hat{n}} G_0}(p) = -\frac{k_\nu^2}{4R_\omega a^3(p)}, \quad \delta\hat{F}(p) = -\frac{a^2(p) + k_\nu^2}{8R_\omega a^6(p)}, \quad \delta\widehat{\partial_{\hat{n}} F}(p) = \frac{2a^2(p) + 3k_\nu^2}{16R_\omega a^5(p)}. \qquad (65)$$

These expressions are slightly different from their counterparts in three dimensions, since their derivation involved the divergence of $p$ which depends on the dimension in which it is calculated. The terms in (65) are inserted into (64), the limit $\nu \to 0$ is performed and the result is multiplied by $2k$ in order to obtain the next term in the asymptotic expansion of $\bar{d}(k)$. The result is

$$\begin{aligned}\bar{d}(k) &= \frac{A}{2\pi}k - \frac{L}{4\pi}\left[1 - \frac{2}{\sqrt{1 + \left(\frac{\kappa}{k}\right)^2}}\right] \\ &\quad + \frac{k}{2\pi^2}\left[\mathrm{Im}\int_\Sigma \frac{\mathrm{d}s_\omega}{R_\omega}\int_{-ik_\nu}^\infty \mathrm{d}a\, \frac{a^2\kappa^2 + k_\nu^2\kappa^2 - a^4 - 2a^2 k_\nu^2}{a^4(a+\kappa)^2\sqrt{a^2 + k_\nu^2}}\right]_{\nu \to 0} + \ldots \qquad (66)\\ &= \frac{A}{2\pi}k - \frac{L}{4\pi}\left[1 - \frac{2}{\sqrt{1 + \left(\frac{\kappa}{k}\right)^2}}\right] + \frac{1}{4\pi k}\frac{1 + 2\left(\frac{\kappa}{k}\right)^2 - \left[1 + \left(\frac{\kappa}{k}\right)^2\right]^{3/2}}{\frac{\kappa}{k}\left[1 + \left(\frac{\kappa}{k}\right)^2\right]^{3/2}}\int_\Sigma \frac{\mathrm{d}s_\omega}{R_\omega} + \ldots\, .\end{aligned}$$

An integration over $k$ gives

$$\bar{N}(k) = \frac{A}{4\pi}k^2 - \frac{L}{4\pi}k\left[1 - 2\left(\sqrt{1 + \left(\frac{\kappa}{k}\right)^2} - \frac{\kappa}{k}\right)\right] + \frac{1}{12\pi}\left[1 - 3\frac{\sqrt{1 + \left(\frac{\kappa}{k}\right)^2} - 1}{\frac{\kappa}{k}\sqrt{1 + \left(\frac{\kappa}{k}\right)^2}}\right]\int_\Sigma \frac{\mathrm{d}s_\omega}{R_\omega} + \ldots\, .$$
$$(67)$$

This is the final result for the leading three asymptotic terms, the area, perimeter and curvature term, for the mean spectral staircase for an arbitrary smooth two-dimensional billiard with mixed boundary conditions. For the determination of the integration constant in the integration of eq. (66) see the dicussion in the section on the circle billiard. Eq. (67) is identical to eq. (25) when applied to a circle. Furthermore, in the limits $\kappa \to \infty$ and $\kappa \to 0$ eq. (67) gives the correct expressions for the case of Dirichlet and Neumann boundary conditions.

## 3.2 The oscillatory part of the spectral density

The oscillatory part of the spectral density for chaotic billiards with mixed boundary conditions (4) is considered in this subsection. We use the scattering approach, in which the billiard is viewed as an obstacle for a scattering problem and use (32) for $d_{osc}(k)$. The scattering matrix is obtained using the Kirchhoff approximation in a way which applies to strictly convex billiards. The approximation assumes that the incoming wave locally sees a straight line at the scattering point. The case of Dirichlet boundary conditions is given in detail in [15]. For mixed boundary conditions the additional phase (9) is again recovered. This phase is slowly varying and in saddle point approximations its value is therefore taken at the saddle point (in section 2.1 this point is justified for the circle).

We consider the scattering solution which consists of an incoming plane wave $\psi_{in}(\vec{r}) = e^{i\vec{k}_i \cdot \vec{r}}$ (with momentum $k_i$ and direction $\theta_i$) and a scattered wave $\psi_{scat}(\vec{r})$. The scattering amplitude $f(\theta_i, \theta_f)$ to an outgoing direction $\theta_f$ is connected to the asymptotic form of the scattered wave at large $r$ by

$$\psi_{scat}(r, \theta_f) \to f(\theta_i, \theta_f) \frac{e^{ikr}}{(2\pi i k r)^{\frac{1}{2}}} \,. \tag{68}$$

The momentum in the outgoing direction is denoted by $\vec{k}_f$, so that $k = |k_i| = |k_f|$. Using Greens theorem the scattering amplitude is expressed as a boundary integral:

$$f(\theta_i, \theta_f) = -\frac{i}{2} \int_\Sigma ds \left[ \partial_{\hat{n}} \psi_{scat}(\vec{r}) + i \left( \vec{k}_f \cdot \hat{n}(\vec{r}) \right) \psi_{scat}(\vec{r}) \right] e^{-i \vec{k}_f \cdot \vec{r}} \,. \tag{69}$$

This expression is exact, but involves the knowledge of both $\psi_{scat}(\vec{r})$ and $\partial_{\hat{n}} \psi_{scat}(\vec{r})$ on the boundary. The boundary condition (4) gives one linear combination of these two quantities, another is found by using a short wave-length approximation.

Given an incoming direction $\theta_i$, the billiards boundary is divided into an illuminated part ($\Sigma_i$) and a shaded part ($\Sigma_s$), dividing the expression for the scattering amplitude (69) into two parts, $f = f_i + f_s$. For the shaded part the approximation is introduced by arguing that the wave function is vanishingly small close to $\Sigma_s$ and hence $\psi(\vec{r}) \cong \partial_{\hat{n}} \psi(\vec{r}) \cong 0$. This results in exactly the same expression as for Dirichlet boundary conditions, giving a semiclassical approximation for the forward scattering peak [15]. The scattering matrix (in the angle representation) is then connected with the illuminated part only.

The short wave-length approximation assumes that the behavior of the scattered wave at a point on the illuminated boundary is determined by specular reflection off that point and that the surface is locally a straight line.

We first consider a wave scattering off a straight wall, with a normal $\hat{n}$ pointing towards the side where the scattering process is taking place. The incoming wave is given, as before, by $\psi_{in}(\vec{r}) = e^{i\vec{k}_i \cdot \vec{r}}$. The scattered wave at any point $\vec{r}$ on the wall and its normal derivative at that point, are expressed in terms of the incoming wave by

$$\psi_{scat}(\vec{r}) = -e^{-i\phi}\psi_{in}(\vec{r}) , \tag{70}$$

$$\partial_{\hat{n}}\psi_{scat}(\vec{r}) = i(\vec{k}_i \cdot \hat{n})e^{-i\phi}\psi_{in}(\vec{r}) . \tag{71}$$

Imposing the boundary conditions at the wall, the additional phase $\phi$ is determined to be

$$\phi = 2\arctan\left(\frac{k}{\kappa}\cos\theta\right) = 2\arctan\left(-\frac{\vec{k}_i \cdot \hat{n}}{\kappa}\right) , \tag{72}$$

where $\theta$ is the angle of incidence and is just the additional phase (9) which appears for mixed boundary conditions. Note that (71) is valid for a wall at any angle or position.

We use the result for the straight wall in (69) to obtain the approximation for the scattering matrix

$$S(\theta_i, \theta_f) = \frac{1}{2\pi}f_i(\theta_i, \theta_f) = \frac{1}{4\pi}\int_{\Sigma_i}ds\,(\vec{k}_i - \vec{k}_f)\cdot\hat{n}(\vec{r})e^{i(\vec{k}_i - \vec{k}_f)\cdot\vec{r} - i\phi(\vec{r})} , \tag{73}$$

where

$$\phi(\vec{r}) = 2\arctan\left(-\frac{\vec{k}_i \cdot \hat{n}(\vec{r})}{\kappa}\right) . \tag{74}$$

The integral (73) is evaluated using a saddle point approximation, where the additional phase $\phi(\vec{r})$ is taken at the saddle point. The saddle point $\vec{r}_0$ for a general convex billiard, is the point of specular reflection for a trajectory with incoming direction $\theta_i$ and outgoing direction $\theta_f$. The scattering matrix is finally given by

$$S(\theta_i, \theta_f) = -\frac{1}{2}\left(\frac{i}{2\pi}\right)^{\frac{1}{2}}\left[(\vec{k}_f - \vec{k}_i)\cdot\hat{n}(\vec{r}_0)R(\vec{r}_0)\right]^{\frac{1}{2}}e^{i(\vec{k}_i - \vec{k}_f)\cdot\vec{r}_0 - i\phi(\vec{r}_0)} , \tag{75}$$

where $R(\vec{r})$ is the radius of curvature and $\phi(\vec{r})$ is given by (74). Note that, since $\vec{r}_0$ is the point of specular reflection, we have $\vec{k}_f \cdot \hat{n}(\vec{r}_0) = -\vec{k}_i \cdot \hat{n}(\vec{r}_0) = k\sin(|\theta_f - \theta_i|/2)$. The scattering matrix for the circle may be directly obtained using $R(\vec{r}_0) = R$ and $\vec{r}_0 = R\hat{n}(\vec{r}_0)$. By transforming to the angular momentum representation, using a saddle point approximation, we recover (29).

The oscillating part of the spectral density, given in the scattering approach by (32), is evaluated using the saddle point method, and the additional phase $\phi$ is taken at the saddle point. The calculations for Dirichlet boundary conditions [15] may then be followed, with the appearance of the additional phases being the only difference. Assuming that all periodic orbits are isolated, we obtain the Gutzwiller sum

$$d_{osc}(k) = \frac{1}{\pi}\sum_{\gamma}\sum_{m=1}^{\infty}L_\gamma\frac{\cos\left[m(kL_\gamma - \nu_\gamma\frac{\pi}{2} - n_\gamma\pi + \Phi_\gamma)\right]}{\sqrt{|2 - \text{Tr}M_\gamma^m|}} , \tag{76}$$

where the sum is over primitive periodic orbits $\gamma$ with repetitions $m$. Each primitive orbit has a length $L_\gamma$ and a monodromy matrix $M_\gamma$. The number of reflections along the orbit is $n_\gamma$ and $\nu_\gamma$ is the maximal number of conjugate points along the orbit (the Maslov index).

The only difference from the formula for Dirichlet boundary conditions is the appearance of the additional phase $\Phi_\gamma$, which is just the sum of the additional phases accumulated along the orbit:

$$\Phi_\gamma = \sum_i \phi_\gamma^i = 2 \sum_i \arctan\left(\frac{k}{\kappa} \cos \theta_\gamma^i\right) , \qquad (77)$$

where $\theta_\gamma^i$ are the angles of incidence at the reflection points along the orbit which are labeled by $i$.

## 3.3 The Spectral Density for the Sinai Billiard

In this subsection we describe very briefly the semiclassical quantization of the two–dimensional Sinai billiard (disc of radius $R$ embedded inside a square of side $L$) with mixed boundary conditions (4) on the disc. For this purpose we follow Berry [8] and use the KKR method [24, 25, 26]. We first unfold the billiard into an infinite lattice and consider a general potential with circular symmetry and range $R$ in the center of each cell ($2R < L$). Using the Green theorem and taking advantage of the translational and circular symmetries, we arrive at a set of linear equations, that have a non–trivial solution (i.e., an eigenvalue of the billiard) if

$$\zeta(k) = \det\left[\delta_{ll'} - \tan\left(\frac{\varphi_l(k)}{2}\right) a_{ll'}(kL)\right] = 0 , \qquad (78)$$

where $a_{ll'}$ are "structure constants" of the lattice that are independent on the potential and $\varphi_l(k)$ are the scattering phase shifts that include all the information about the potential. For mixed boundary conditions the phase shifts are given by

$$S_l = e^{-i\varphi_l} = -\frac{\kappa H_l^-(kR) - k H_l^{-'}(kR)}{\kappa H_l^+(kR) - k H_l^{+'}(kR)} , \qquad (79)$$

which is different from eq. (27) for the circle, due to $\partial_{\hat{n}} = -\partial_r$ in the Sinai billiard case. Semiclassically we have

$$\varphi_l(k) = \varphi_l^D(k) - \phi_l(k) , \qquad (80)$$

where $\varphi_l^D(k)$ are the phase shifts for Dirichlet boundary conditions on the disc and $\phi_l = 2 \arctan\left[(k/\kappa) \cos \theta_l\right]$. $\theta_l$ is the angle of incidence to the disc of a classical particle having angular momentum $l$.

To calculate the expression for the spectral density $d(k)$ one considers $(-1/\pi) \log \zeta(k)$ ($\zeta(k)$ can be made real [8]). Then one uses the matrix relation log det = Tr log and formally expands the logarithm in a Taylor series. To get the semiclassical approximation of $d(k)$, Poisson summation is used and the resulting integrals are evaluated by using saddle point approximation term by term. (The detailed and rather long derivation is one of the main subjects of [8].) Substituting $\varphi_l$ in (78) and performing these steps, we immediately get a semiclassical expression for the Sinai billiard with mixed boundary conditions that contains no expansions in powers of $\kappa$. Thus "$\kappa$ is resummed to all orders" by the KKR determinant and is included only through the phase shift $\phi_l$. Moreover, due to the slow variation of $\phi_l$ as explained above (see subsection 2.1), the modifications are particularly simple and result in

$$d(k;\kappa) = \bar{d}(k;\kappa) + d_{osc}(k;\kappa) = \frac{k(L^2 - \pi R^2)}{2\pi} - \frac{L}{\pi} - \frac{R}{2}\left[1 - \frac{2}{\sqrt{1 + \left(\frac{\kappa}{k}\right)^2}}\right] + d_{bb}(k)$$

$$+ \sum_{\gamma}\sum_{m=1}^{\infty} A_{\gamma,m} \cos\left[mkL_\gamma - mn_\gamma\pi + m\sum_{i=1}^{n_\gamma^c} 2\arctan\left(\frac{k}{\kappa}\cos\theta_\gamma^i\right)\right], \qquad (81)$$

where $A_{\gamma,m}$ are the semiclassical amplitudes that appear in (76) and $n_\gamma$ are the number of collisions of the primitive periodic orbit $\gamma$ with the billiard walls. Note the fact that for the Sinai billiard the Maslov index $\nu_\gamma$ is always 0. The inner sum is over collisions with the disc ($\theta_\gamma^i$ are angles of incidence) and contains the modifications to $d_{osc}(E)$ due to mixed boundary conditions. The $\kappa$–independent term $d_{bb}(k)$ contains the non–generic contributions of the "bouncing ball" orbits [8, 9, 27] and will be further discussed below (section 4). It is interesting to note, that although in eq. (80) the sign of the additional phase shift $\phi_l$ is opposite to the that of the circle (and more generally the convex) billiard, the final results for both $\bar{d}(E)$ and $d_{osc}(E)$ agree in signs with the previous cases, which emphasizes the generality of the modifications implied in eq. (81).

## 4 Application of Mixed Boundary Conditions to the Spectral Diagnostics of the Sinai Billiard

The spectral density of the Sinai billiard, in either two or three dimensions, and in the presence of mixed boundary conditions on the inscribed scatterer (either disc or sphere), can be written as

$$d(k;b,\alpha) = \bar{d}(k;b,\alpha) + d_{osc}(k;b,\alpha), \qquad (82)$$

$$d_{osc}(k;b,\alpha) = d_{gen}(k;b,\alpha) + d_{bb}(k) + d_{tan}(k;b,\alpha) \qquad (83)$$

where $\bar{d}$ is the smooth density, $d_{gen}$ comes from contributions of generic, isolated and unstable periodic orbits (see (81)), $d_{bb}$ represents the contributions of the non–generic "bouncing ball" (BB) families of marginally stable orbits that do not collide with the scatterer [9, 8, 27], and $d_{tan}$ contains the non–generic contributions due to periodic orbits which are tangential to the scatterer (see figure 3). When one is interested in the generic part of the spectral density, one must find a reliable and accurate method for the elimination of the BB and tangent contributions. This is not an easy task, since the contribution of the leading–order BB families are $O(k^{1/2})$ stronger than that of an isolated periodic orbit in the two–dimensional case [9], and $O(k^1)$ stronger in three dimensions. In the semiclassical limit this may amount to a very large factor, so that to identify the contributions of a single, generic orbit one must subtract the BB contributions in an accurate and reliable way. This problem is especially acute when one considers the cosine transform of the oscillating part of the spectral density – the length spectrum:

$$D(x;k_{max};b,\alpha) \equiv \frac{1}{k_{max}}\int_0^{k_{max}} dk\, d_{osc}(k;b,\alpha)\cos(kx). \qquad (84)$$

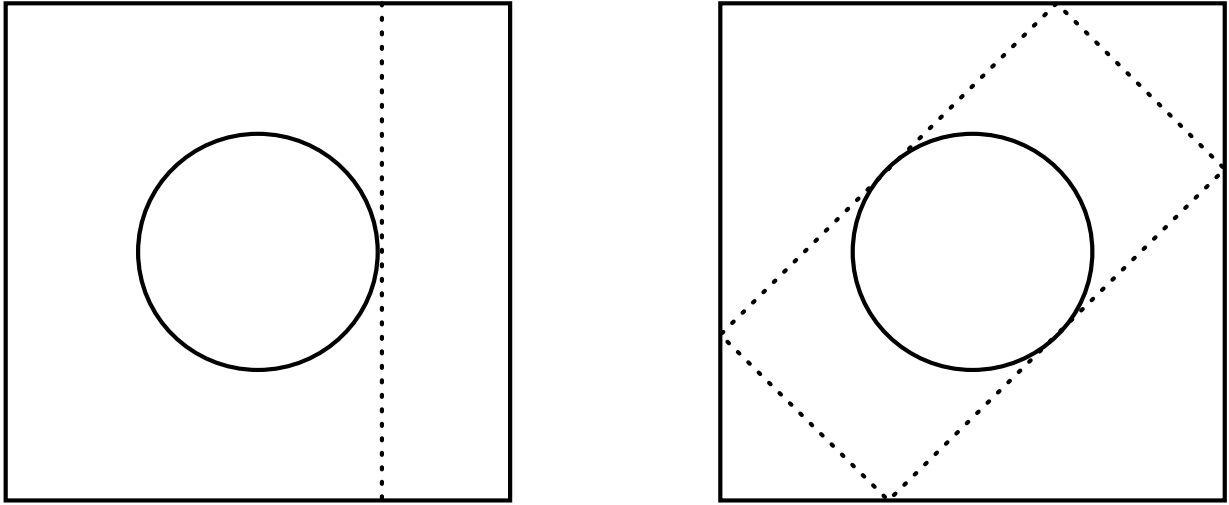

Figure 3: Tangent periodic orbits in the two–dimensional Sinai billiard.

Here, due to the finite range of $k$ values, the length spectrum is composed of peaks with a finite width, and the contribution of the BB families might overlap and obscure the peaks due to unstable, isolated periodic orbits. This problem is very well demonstrated in the top part of figure 4. At the length $x = 2$ there exist a BB family and an isolated periodic orbit (a two fold repetition of the shortest orbit of length 1). The latter has an amplitude which is smaller even than that of the peak at $x = 1$.

In order to eliminate the non–generic contributions, it is natural to seek analytic expressions, and to subtract them from the spectral density. Indeed, for the two–dimensional stadium billiard the contribution of the (single) BB family were calculated by Sieber et al. [9], and were successfully applied in the spectral analysis. These calculations become, however, much more intricate for the Sinai billiard, in particular for three dimensions. Instead of a single BB family as in the stadium, the Sinai billiard admits a rich variety of BB families, especially in three dimensions [28]. Moreover, in three dimensions the total (geometrical) measure of each family of BB is difficult to evaluate. Tangent orbits (which are not encountered in the stadium billiard) require a special treatment, which is not yet available.

Facing all these difficulties, we were forced to introduce a new method for the elimination of the BB and tangent orbits. It makes use of the extra freedom which is gained by the variation of the boundary conditions. As a matter of fact our interest in the semiclassical theory of billiards with mixed boundary conditions stemmed originally from this particular application.

The main idea of this method is to look at differences of spectral densities obtained with boundary conditions which differs only on the scatterer. Since the BB orbits do not collide with the scatterer, they are insensitive to changes of the boundary conditions on it. Therefore, their contributions will be eliminated in the difference. We shall discuss here two convenient schemes. In the first, we shall subtract the spectrum with Dirichlet boundary condition on the scatterer from the spectrum with the Neumann condition on the scatterer. The difference will be called the N-D spectrum.

Taking into account the explicit expressions for $d_{osc}$ for Neumann and Dirichlet boundary conditions we get from eq. (81)

$$d_{Neu}(k) - d_{Dir}(k) = \text{(smooth part)} + 2 {\sum_{\gamma,m}}' A_{\gamma,m} \cos\left(mkL_\gamma - m\, n_\gamma^s \pi\right) + d'_{tan}, \qquad (85)$$

where $n_\gamma^s$ is the number of collisions with the straight boundaries, and the prime over the sum indicates that only periodic orbits with an odd number of collisions with the scatterer should be considered. $d'_{tan}$ is the residual contribution of the tangent orbits. The BB contribution $d_{bb}(k)$ is completely eliminated by the subtraction. This result is valid in the semiclassical approximation, and its applicability in practice should be tested. This was done by analyzing numerically the two–dimensional quarter of a Sinai billiard with $L = 2, R = 0.5$. Our numerical database consists of the lowest 5700 energy levels [29] in the $k$–range $0 < k_n < 300$, for both Dirichlet and Neumann boundary conditions on the disc.

In figure 4 (top) we present the two length spectra for Neumann and Dirichlet boundary conditions on the disc and one clearly observes the large contributions of the BB families, which have lengths of 2 and $2\sqrt{2}$ [8]. The change of sign between Dirichlet and Neumann boundary conditions due to odd number of bounces on the disc is evident, e.g. for the shortest orbit of length 1 which bounces only once from the disc. There are also peaks that do not change sign, as expected. The N-D spectrum is shown in figure 4 (lower part). The peaks corresponding to the BB are greatly diminished, and generic periodic orbits can be distinguished more easily. There are, however, some drawbacks to the N-D analysis. The first one is that it eliminates half of the generic contributions to the spectral density, namely those orbits that bounce an even number of times from the scatterer (see eq. (85)). This can be easily rectified by comparing spectra with other boundary conditions on the scatterer rather than the special N-D analysis. A more serious drawback is due to the contributions of tangent orbits. They have the same length as the BB, and their residual contribution can be clearly seen in figure 4 for $x = 2, 2\sqrt{2}$. It was found numerically to be of the same order in $k$ as for isolated orbits. This gives an indication, that in three dimensions such tangent orbits, that form 1–parameter families, are likely to give contributions of order $k^{1/2}$, which can obscure a significant part of the generic contributions.

The second, and more powerful comparison method consists of taking the derivative of $d(k; b, \alpha)$ with respect to $\alpha$ at $\alpha = 0$. Using eqs. (9, 81, 83) we get

$$\begin{aligned}
\left.\frac{\partial}{\partial \alpha} d(k; b, \alpha)\right|_{\alpha=0} &= \sum_{n=1}^{\infty} \delta'(k - k_n)\left(\frac{dk_n}{d\alpha}\right)_{\alpha=0} \\
&= \left.\frac{\partial}{\partial \alpha} \bar{d}(k; b, \alpha)\right|_{\alpha=0} + \sum_{\gamma,m} A_{\gamma,m} \sin(mkL_\gamma - m n_\gamma \pi)\left[-2m\frac{k}{b}\sum_{i=1}^{n_\gamma^c} \cos\theta_\gamma^i\right] \\
&\quad + \left.\frac{\partial}{\partial \alpha} d_{tan}(k; b, \alpha)\right|_{\alpha=0}.
\end{aligned} \qquad (86)$$

The term $d_{bb}$ dropped trivially, and the oscillatory contributions are now multiplied by the new prefactors $\sum_{i=1}^{n_\gamma^c} \cos\theta_\gamma^i$. These prefactors vanish (only) for tangent orbits ($\theta_\gamma^i = \pi/2, i = 1, \ldots, n_\gamma^c$). If we now assume that the tangent orbit contributions behave similarly to those of the periodic orbits, it will follow that their contributions will also be eliminated. Hence, the $\alpha$ derivative of the spectral density consists of generic contributions, exclusively.

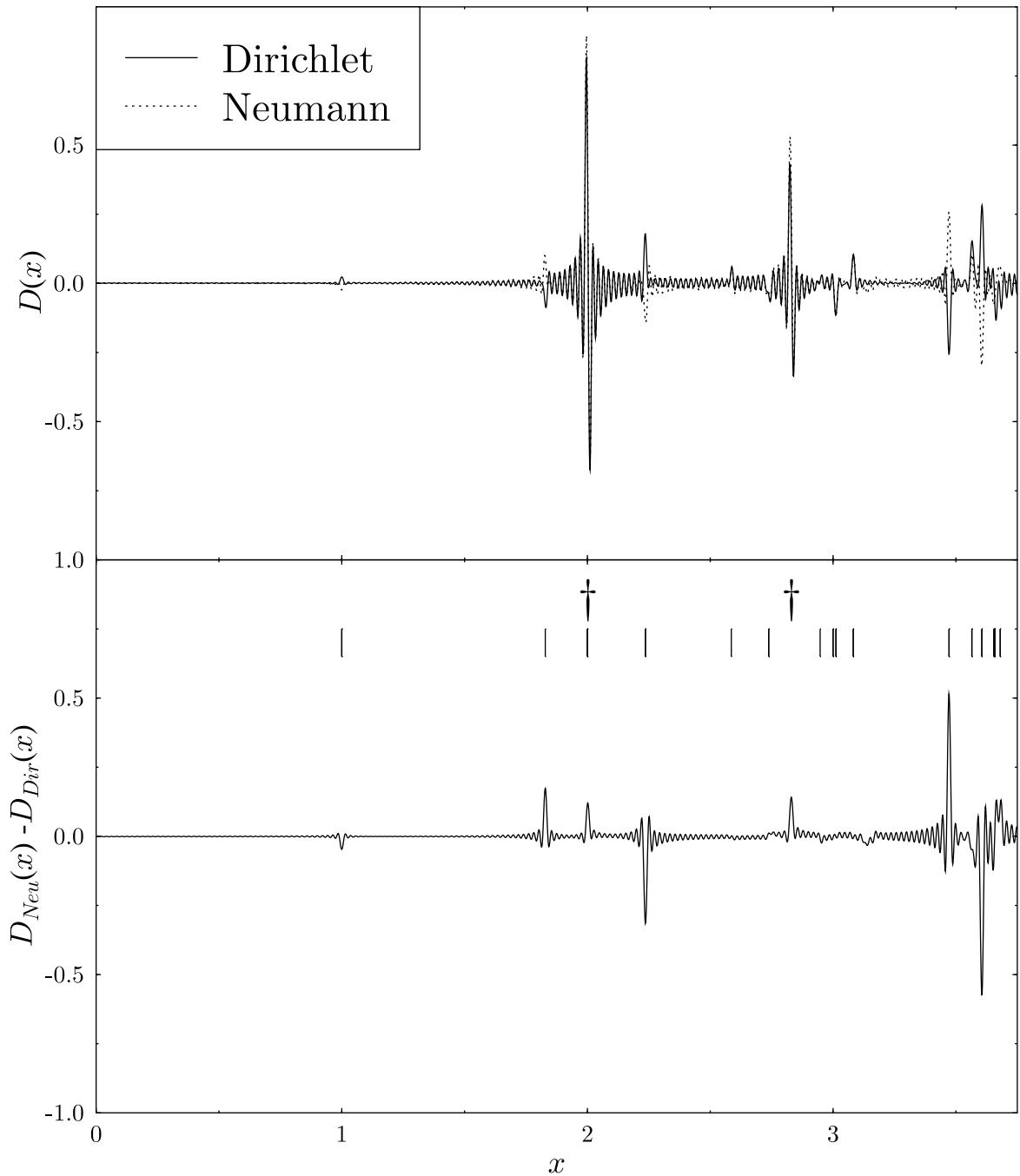

Figure 4: Top: length spectra of quarter of the two–dimensional Sinai billiard, $L = 2, R = 0.5$, with Neumann and Dirichlet boundary conditions on the disc. Bottom: difference between Neumann and Dirichlet length spectra. Isolated periodic orbits are indicated by vertical lines and bouncing ball manifolds are indicated by daggers.

To examine (86) we use instead of the plain length spectrum (84) a Gaussian modified one with the kernel

$$f(k, x; k_0, \sigma) = \frac{1}{\sqrt{2\pi\sigma^2}} \exp\left(-\frac{(k-k_0)^2}{2\sigma^2}\right) \cos(kx) \,. \tag{87}$$

The parameters $k_0$ and $\sigma$ were chosen, such that for the calculated spectra, the integration limits can be safely taken to $\pm\infty$, thus avoiding undesired boundary effects. Integrating both sides of (86) with this kernel, one gets

$$\sum_{n=1}^{\infty} -\frac{\exp\left(-\frac{(k_n-k_0)^2}{2\sigma^2}\right)}{\sqrt{2\pi\sigma^2}} \left[\frac{k_n-k_0}{\sigma^2}\cos(k_n x) + x\sin(k_n x)\right] \left(\frac{\mathrm{d}k_n}{\mathrm{d}\alpha}\right)_{\alpha=0} =$$
$$e^{-\frac{x^2\sigma^2}{2}} \left\{\frac{R}{4b}\left[k_0\cos(k_0 x) - x\sigma^2\sin(k_0 x)\right] + \left(\frac{1}{8} - \frac{1}{2\pi}\right)\frac{1}{b}\cos(k_0 x)\right\} +$$
$$\sum_{\gamma,m} \tilde{A}_{\gamma,m} \left\{e^{-\frac{(mL_\gamma-x)^2\sigma^2}{2}}\left[(mL_\gamma-x)\sigma^2\cos(k_0(mL_\gamma-x)) + k_0\sin(k_0(mL_\gamma-x))\right] +\right.$$
$$\left. e^{-\frac{(mL_\gamma+x)^2\sigma^2}{2}}\left[(mL_\gamma+x)\sigma^2\cos(k_0(mL_\gamma+x)) + k_0\sin(k_0(mL_\gamma+x))\right]\right\} \,, \tag{88}$$

where

$$\tilde{A}_{\gamma,m} = -\frac{A_{\gamma,m}}{b}(-1)^{\frac{mn_\gamma^c}{2}} m \sum_{i=1}^{n_\gamma^c} \cos\theta_\gamma^i \,. \tag{89}$$

Here we used the explicit expression for the smooth spectral density of the two–dimensional quarter Sinai billiard including the corner and curvature terms. To check eq. (88), we calculated numerically the derivatives of energy levels with respect to $\alpha$ by using finite differences for $\alpha = 0$ and $\alpha = 0.0005\,\pi/2$. In both cases $b = 200$ and $0 < k_n < 300$. In figure 5 we present the quantum results together with the semiclassical predictions of eq. (86) which consists of contributions due to generic isolated and unstable periodic orbits (and their repetitions) only. The agreement between the quantum and the semiclassical results is good throughout the entire length interval (but for a few deviations which are due to special effects which will be discussed latter). This demonstrates the power of the derivative method, and the general success of the semiclassical theory.

A more detailed examination of figure 5 reveals the following details. At the lengths 2 and $2\sqrt{2}$ which correspond to the shortest BB families and their limiting tangent orbits, we can observe very small structures. Comparing them to the corresponding peaks in the N-D spectrum, we see that the tangent orbits behave as expected. The very small peaks that nevertheless exist there, are attributed to higher–order corrections to the semiclassical approximation and to the small contribution of the second repetition of the shortest periodic orbit of length 1. Other BB families cannot be separately examined for the given $k$ range due to the clustering of periodic orbits (the total number of periodic orbits grows exponentially with their length).

It is important to note that the quality of the semiclassical theory is uniform throughout the length domain, even at longer lengths, where individual periodic orbits cannot be resolved. This is a very demanding test for the semiclassical theory since it involves the interference of dozens of periodic orbits per wavelength. To get such a good reconstruction of the exact length spectrum requires a very accurate evaluation of the individual amplitudes and phases.

The largest discrepancy between the numerical data and the semiclassical theory is seen in the vicinity of $x = 6.21$. At this length there exists a triplet of unstable periodic orbits which are very close to tangency with the disc. It would be eliminated by pruning if the radius of the disc was increased beyond $R = 0.555$. Other points in the length spectrum where deviations of the semiclassical theory from the numerical length spectrum can be observed are also close to bifurcations of this type with two or three periodic orbits becoming simultaneously tangent to the disc. Therefore we suspect that the deviations have their origin in the failure of the semiclassical theory in its present form in the vicinity of tangent orbits.

The method presented above was recently applied in the analysis of the 3-D Sinai billiard. There, the length spectrum is entirely overwhelmed by BB contributions. One can study the contributions of isolated periodic orbits only after the spectrum is analyzed by the methods explained above. The detailed analysis of the 3-D Sinai billiard will be presented elsewhere. Finally, we would like to note that the method presented here can be applied in many variations, whenever a particular set of orbits is of particular interest, and one wishes to suppress all other contributions.

## 5 Parametric Spectral Statistics

Random matrix theory has been successful in describing universal properties in the spectra of a large variety of *single* systems. These include classically chaotic systems as well as disordered mesoscopic systems. There is now an increased interest in considering not only single isolated systems but also systems that depend on an external parameter and in describing correlation functions in this parameter and statistics that depend on the sensitivity of the energy levels to the change of the parameter [10, 11, 12, 13].

Two very often considered quantities are the velocity and the curvature distributions, i.e. the distributions of the first and second derivatives of the levels with respect to the parameter. For these quantities several theoretical results exist. The velocity distribution $P(v)$ is a Gaussian for random matrix ensembles. It has been shown by Gaspard et al. [10] that the asymptotic distribution for large curvatures is universal. Subsequently, this behavior was verified numerically for several systems [30, 31, 12]. Formulae for the *complete* curvature distributions for the GUE, GOE and GSE of random matrix theory were conjectured by Zakrzewski and Delande [12] based on detailed numerical examinations. In addition, it was suggested that deviations from these distributions in chaotic systems could be taken as a measure of the degree of scarring in this system [31, 12]. It has been proved recently that the conjectured distributions are indeed the exact distributions for the random matrix ensembles [32, 33].

In this section we examine the velocity and curvature distributions for the Sinai billiard with mixed boundary conditions on the dispersing arc. As has been mentioned in the introduction, it is a special feature of this parameter dependence that it restricts only to the quantum system, whereas the corresponding classical system is parameter-independent.

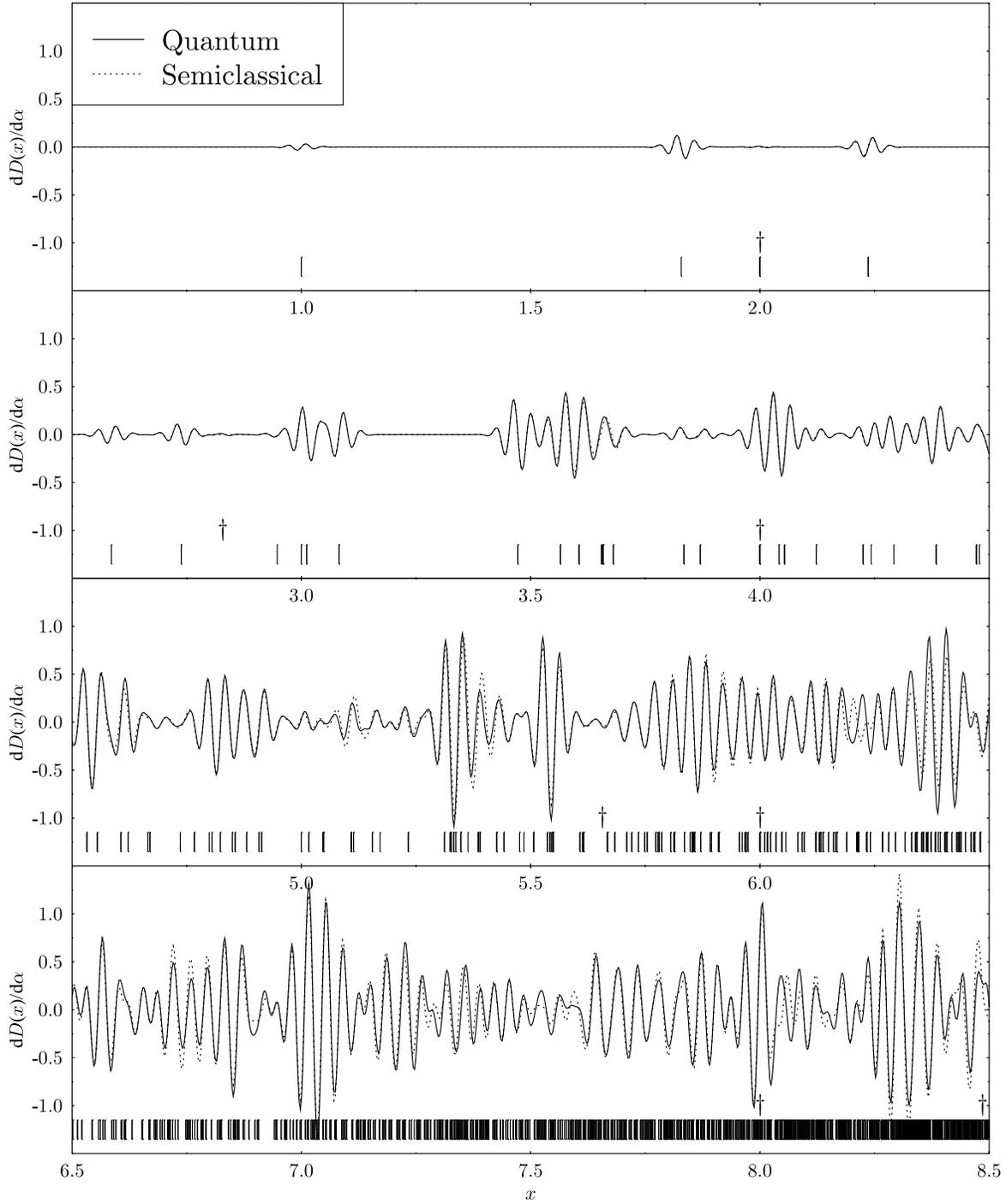

Figure 5: The derivative of the length spectrum of the quarter Sinai billiard, $L = 2, R = 0.5$, with respect to $\alpha$ at $\alpha = 0$. Lengths of periodic orbits are indicated by vertical lines, lengths of BB families are indicated by daggers.

As parameter we take the quantity $\alpha$ of eq. (5) with fixed $b = 200$. The velocities are

$$v_n = \frac{\partial x_n}{\partial \alpha}\sqrt{\beta(x_n)} \quad , \quad \beta(x) = \left\langle \left(\frac{\partial x_n}{\partial \alpha}\right)^2 \right\rangle_x^{-1} , \qquad (90)$$

where $x_n(\alpha) = \bar{N}(k_n(\alpha); b, \alpha)$ are the unfolded energy levels. We scaled the first derivatives $\partial x_n/\partial \alpha$ by their local standard deviation, since this standard deviation is not constant with increasing energy. The brackets $<\cdot>_x$ denote an average over the levels $x_n$ in a local interval around the value $x$. The scaled curvatures are defined as

$$c_n = \frac{\partial^2 x_n}{\partial \alpha^2}\frac{\beta(x_n)}{\nu\pi} , \qquad (91)$$

where $\nu$ classifies the corresponding random matrix ensemble. It is equal to one for GOE, which is appropriate for systems with time reversal symmetry. The distribution of curvatures for GOE is given by

$$P(c) = \frac{1}{2}\frac{1}{(1+c^2)^{3/2}} . \qquad (92)$$

We compare the numerical distributions for the Sinai billiard to the random matrix results and discuss the deviations that occur, especially in the context of the scarring of wave functions in the system. The calculations were done for the quarter Sinai billiard which has two symmetry classes, odd and even. The eigenfunctions of these symmetry classes are solutions of the Helmholtz equation for the fundamental region of the Sinai billiard, which is drawn in figure 7 with thick lines, with Dirichlet (odd) or Neumann (even) boundary conditions along the line $y = x$, mixed boundary conditions along the circular part of the boundary and Dirichlet boundary conditions along the remaining two straight sections of the boundary. As in the previous section $R = 0.5$ and $L = 2$ and for these parameter values there exist two families of bouncing ball orbits. For the numerical analysis the energy levels with a wave number in the range $100 < k < 300$ were used. This corresponds to approximately 2500 levels in each symmetry class. The statistical distributions were evaluated by averaging over the two symmetry classes.

## 5.1 Non-Generic Features in the Parameter Dependent Energy Spectrum

Figure 6 is a plot of the wave numbers $k_n$ of a part of the spectrum as a function of the parameter $\alpha$ for the two symmetry classes of the quarter Sinai billiard. There is a mean decrease of the levels with increasing $\alpha$ which is due to the $\alpha$-dependence of the mean number of energy levels $\bar{N}(k)$. A striking feature in figure 6 is that one can distinguish apparent horizontal lines. These lines consist of almost straight pieces which are interrupted by avoided level crossings. As will be argued below these lines can be attributed to the existence of the families of bouncing ball orbits. We discuss these lines in more detail since they lead to deviations in the velocity and curvature distributions from the random matrix theory. Similar apparent lines have also been found in various other systems (see e. g. [34]) and also in experiments [35].

The lines are regularly spaced and they can be divided into two families. The first family appears at the same positions in the spectra of both symmetry classes. The lines actually appear as double lines where the upper line is less pronounced than the lower line. The positions are roughly given by $k = n\pi$, where $n$ is a natural number (dark triangles in figure 6). In the case of the second family of lines there is a difference between the two symmetry classes. One can distinguish horizontal lines at the approximate positions $k = n\pi/\sqrt{2}$, but they appear only for odd integers $n$ in the odd symmetry class and for even integers $n$ in the even symmetry class (empty triangles in figure 6). Although we did not calculate wave functions, we strongly suspect that the horizontal lines correspond to wave functions that are mainly concentrated (scarred) along one of the two families of bouncing ball orbits and that these states are not strongly influenced by changing the boundary conditions on the circular part of the boundary. Under these assumptions the actual position of the lines and the difference between the two symmetry classes can be explained by a very simple model, namely by approximating the wave functions by the superposition of two eigenfunctions of appropriate rectangular billiards.

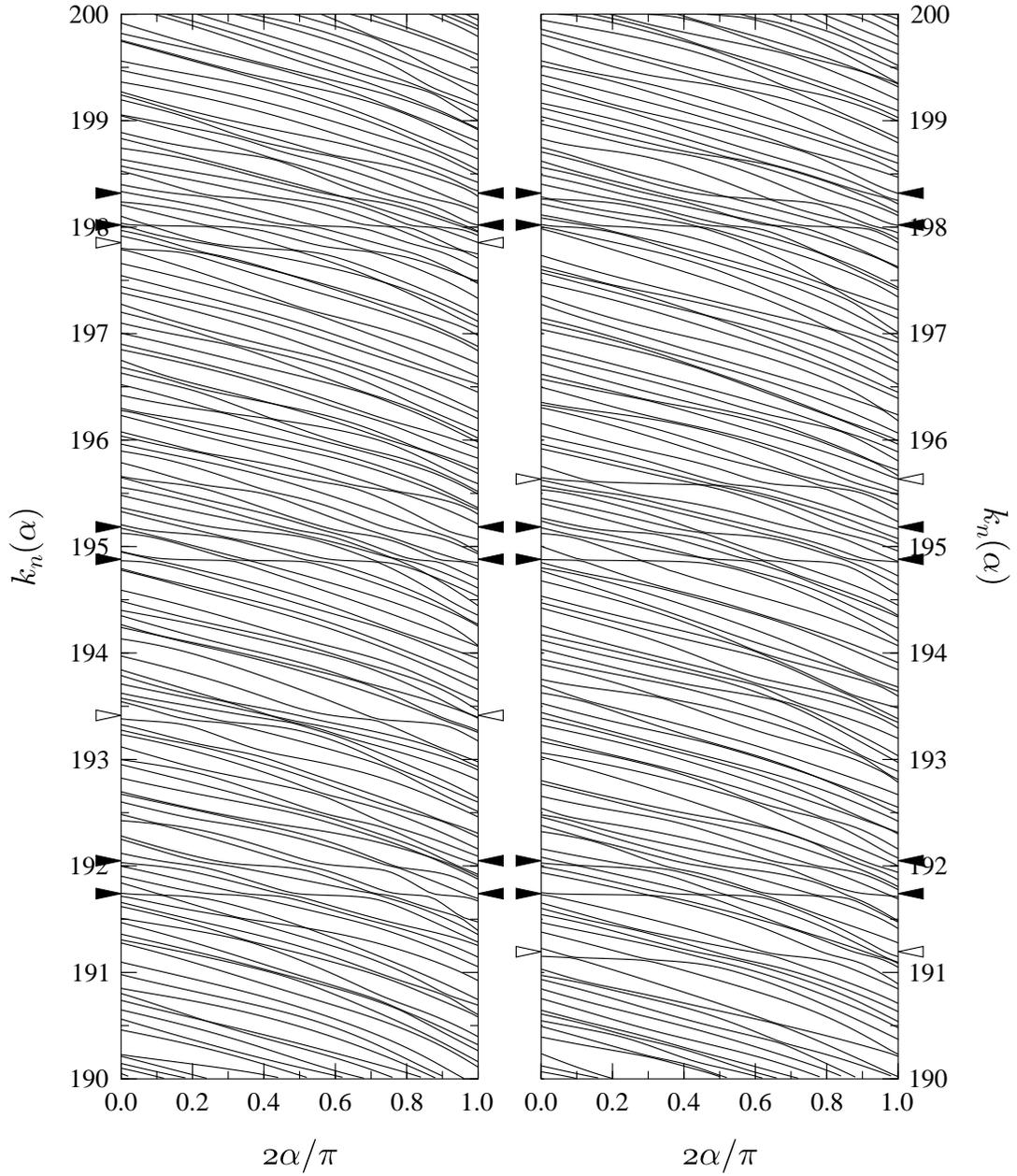

Figure 6: The wave numbers $k_n(\alpha)$ in the range $190 < k_n < 200$ as functions of the parameter $\alpha$ for the desymmetrized Sinai billiard, for odd (left figure) and even (right figure) symmetry. The triangles indicate eigenvalues of two rectangular billiards (dark triangles – first family, empty triangles – second family), as discussed in the text.

We consider first the position of the lines in the vicinity of $\alpha = 0$. Later on, we shall show that the lines are not strictly horizontal, but that they drop slightly as $\alpha$ approaches $\pi/2$. We start with the first family of lines. It is convenient to discuss them in the quarter Sinai billiard instead of the fundamental region. We assume that the wave functions that correspond to these lines are scarred along the first family of bouncing ball orbits. These are orbits that are parallel to the $x$-axis or the $y$-axis in the coordinate system shown in figure 7, i.e. they occupy the region of the upper shaded rectangle in figure 7 and the region of another rectangle that is obtained by reflecting the first rectangle on the line $y = x$. We then express the wave functions by a linear superposition of a function $f(x, y)$ and its mirror image $f(y, x)$, where $f(x, y)$ is concentrated along the first of the rectangles only. The wave functions of the two symmetry classes are then given by $f(x, y) \mp f(y, x)$. This decomposition in terms of $f(x, y)$ and $f(y, x)$ is of course only possible within an approximation since it implies that both symmetry classes have the same eigenvalue.

The fact that the eigenvalues of the wave functions are only little influenced by the change of the boundary conditions on the circular part of the boundary points to a small excitation of the function $f(x, y)$ in $x$-direction and a high excitation in $y$-direction. For that reason, we determined its eigenvalue by an adiabatic approximation in which the high excitation in $y$-direction and the low excitation in $x$-direction are separated. In the energy range considered, the results differ only little from the eigenvalues of the rectangular billiard in figure 7 with Dirichlet boundary conditions. For simplicity we therefore restrict the discussion to the eigenvalues of this rectangular billiard. The wave numbers of its eigenvalues are given by $k_{n,m} = \pi\sqrt{n^2/a^2 + m^2/b^2}$, where $n$ and $m$ are integers and $a = 1$ and $b = 1 - R$ are the lengths of its sides. The wave numbers $k_{n,m}$ provide a more detailed description of the positions of the lines than the rough estimate $k = n\pi$. The values of $k_{n,m}$ are marked in figure 6 for $n = 61$ to $63$ and $m = 1$ and $2$ by black triangles. As one can see, they agree well with the position of the horizontal lines. Scarred wave functions that correspond to higher values of $m$ might also exist, but they are more sensitive to a change of the parameter $\alpha$ and therefore the corresponding lines cannot be clearly distinguished in the spectrum.

For the consideration of the second family of lines we introduce new coordinates $u = (x - y)/2$ and $v = (x + y)/2$, and consider a region that is bounded by the lines $u = 0$, $u = 1$, $v = 0$ and $v = 1$ and two circular arcs as shown in figure 7. Let us denote this region by $D$. It contains four copies of the fundamental region of the billiard which is indicated by thick lines. The second family of bouncing ball orbits runs along lines of constant $u$ or constant $v$. In order that a wave function in the domain $D$ is also an eigenfunction of the two considered symmetry classes of the fundamental region, it has to satisfy Dirichlet or Neumann boundary conditions, respectively, on the straight sections of the boundary of $D$ and additionally it has to vanish on the diagonals $v = u$ and $v = 1 - u$.

We consider now a wave function which is mainly concentrated along the family of bouncing ball orbits that is parallel to the $v$-axis. This region is indicated by the tilted rectangular region in figure 7. Again we try to approximate the energy of this scarred wave function by an eigenvalue of the rectangular billiard. Depending on the considered symmetry class, the eigenfunction $f(u,v)$ of the rectangular billiard has to satisfy Dirichlet or Neumann boundary conditions on its short sides and Dirichlet boundary conditions on its long sides (we consider the case $\alpha = 0$). From $f(u,v)$ one can obtain a wave function $\psi(u,v)$ in the fundamental region, i.e. one which vanishes on the lines $v = u$ and $v = 1 - u$, by defining $\psi(u,v) = f(u,v) - f(v,u)$ and requiring that $f(1/2 + z, 1/2 - z) = f(1/2 - z, 1/2 + z)$. A consequence of the last condition is that the function $f(u,v)$ must have the same parity with respect to reflection of its first coordinate at $u = 1/2$ as with respect to reflection of the second coordinate at $v = 1/2$.

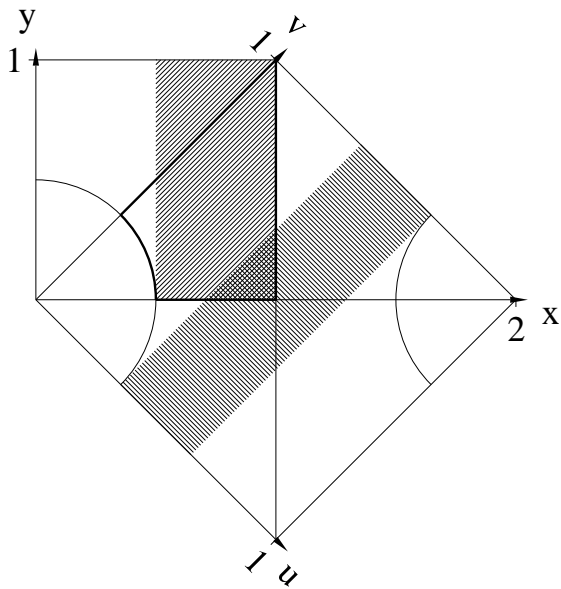

Figure 7: The $(x,y)$- and $(u,v)$-coordinate systems for a description of the two families of bouncing ball orbits in the quarter Sinai billiard with $R = 0.5$ and $L = 2$. The fundamental region of the Sinai billiard is drawn with thick lines.

We now consider the odd symmetry class of the quarter Sinai billiard. For this case the function $f(u,v)$ is given by $f(u,v) = c_{n,m} \sin[\pi n v] \sin[\pi m (u - r)/(1 - 2r)]$, where $c_{n,m}$ is a normalization constant and $r = R/\sqrt{2}$ is the radius of the circle in the $(u,v)$-coordinates. The corresponding wave numbers are $k_{n,m} = \pi\sqrt{n^2/a^2 + m^2/b^2}/\sqrt{2}$ where $a = 1$ and $b = (1 - 2r)$ and the factor $\sqrt{2}$ has its origin in the Jacobian determinant of the coordinate transformation. The symmetry condition that $f(u, v)$ has to satisfy restricts the possible values of $m$ and $n$. It is satisfied if either $m$ and $n$ are both even or if they are both odd. As a consequence, the lowest excitation $m = 1$ in $u$-direction is only possible, when $n$ is an odd number. In the left graph of figure 6 the wave numbers $k_{n,m}$ are indicated by empty triangles for the values $n = 87$ and $89$ and $m = 1$ and one can see close to these positions horizontal lines in the spectrum. The positions of the lines are actually slightly below the values of $k_{n,m}$, possibly because the wave functions spread slightly beyond the rectangular region. The lines are less pronounced than the lines of the first family and already for the cases $m = 2$, which correspond to even values of $n$, there are no clear lines in the spectrum.

For the even symmetry class of the quarter Sinai billiard, the function $f(u, v)$ is given by $f(u, v) = c_{n,m} \cos[\pi n v] \sin[\pi m(u - r)/(1 - 2r)]$. Now the symmetry condition for $f(u, v)$ requires that $m$ is even if $n$ is odd and vice versa. In the right graph of figure 6 the lines of the second family for $m = 1$ have the same qualitative properties as in the left graph, with the difference that they now appear for even values of $n$. The empty triangles in the right graph mark the values of $k_{n,m}$ for the values $n = 86$ and $88$ and $m = 1$.

Up to now, we considered only the position of the horizontal lines in the spectrum near the parameter value $\alpha = 0$. We now discuss the fact that the lines are not strictly horizontal but that they drop slightly when $\alpha$ is increased. In order to describe the lines more accurately over the whole $\alpha$-range we again use the simple model of the two rectangular billiards in figure 7. This time we impose mixed boundary conditions on those sides of the two rectangles that are tangent to the circular arcs in figure 7. The resulting wave numbers $k_{n,m}(\alpha)$ (for the appropriate values of $n$ and $m$) have the following properties. They are monotonously decreasing functions of $\alpha$, yet the decrease is very slight except near $\alpha = \pi/2$, where it becomes stronger. At $\alpha = \pi/2$ the lines have a distinctive negative curvature and negative slope. The model of the rectangular billiard is not good enough to give an exact description of the actual position of the lines in the spectrum, but the qualitative behavior of the functions $k_{n,m}(\alpha)$ that is described above can also be seen in the spectrum.

## 5.2 The Velocity and Curvature Distributions

We now come back to the distribution of velocities and curvatures and consider first the results for the parameter value $\alpha = 0$. The velocity distribution is shown in figure 8 and is compared to a Gaussian of the same mean value and standard deviation. One notices that the distribution is not well approximated by the Gaussian curve. It is asymmetric and it has an excess of large velocities. This deviation from the random matrix result can be explained by the presence of the horizontal lines in figure 6 (see [12]). When the spectrum is unfolded, these lines go over into lines that have a large slope (except near $\alpha = \pi/2$ as will be discussed later). They thus contribute to large positive velocities. Similar results were obtained for the hydrogen atom in a strong magnetic field, a system which has a large number of scarred states [12].

The curvature distribution that is shown in figure 9 is likewise not in agreement with the random matrix result. There is a large excess of small curvatures in comparison with the distribution of eq. (92). These results are again in qualitative agreements with results for the hydrogen atom in a magnetic field and also with results for the stadium billiard where the parameter is the length of the straight section of the boundary [31, 12].

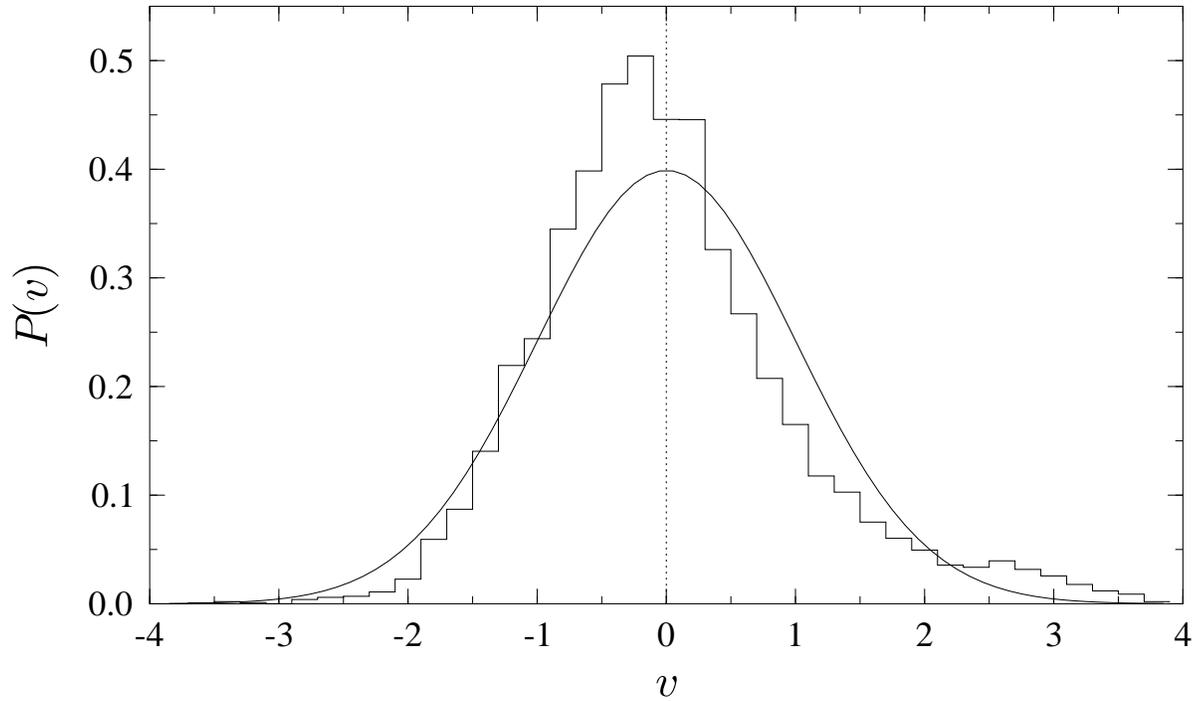

Figure 8: The velocity distribution $P(v)$ at $\alpha = 0$ in comparison with a Gaussian.

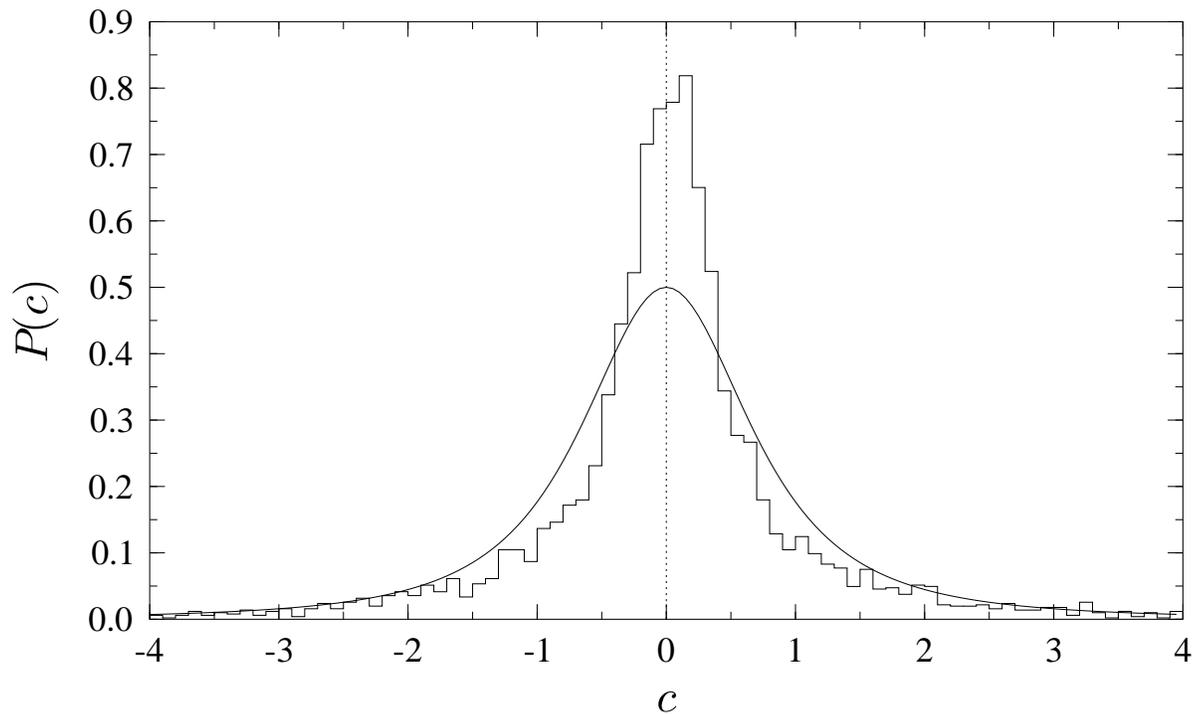

Figure 9: The curvature distribution $P(c)$ at $\alpha = 0$ in comparison with the GOE-distribution of eq. (92).

The high peak of the distribution at small curvatures can be explained by two effects. The horizontal lines in figure 6 contain long pieces where the curvature is almost zero. This is still the case after unfolding the levels and these parts of the spectrum thus contribute to an accumulation of small curvatures. Secondly, as has been noted above they are also responsible for an excess of large velocities. This excess leads to an increase of the variance of the velocity distribution $\beta^{(-1)}$ and, by definition (91), to a decrease of the curvatures. Because of these effects it has been suggested that the excess of small curvatures can be taken as measure for the degree of scarring in a system [31, 12]. We will discuss this point in more detail when considering results for other values of the parameter $\alpha$.

It is possible that the small avoided level crossings along the horizontal lines lead also to a non–generic curvature distribution at large curvatures [12], but the number of levels has not been large enough in order to check this in detail.

In order to see whether the discrepancy between the calculated distributions and the random matrix result is only due to levels in the vicinity of the horizontal lines, we disregarded these levels and determined the distributions with the remaining levels. The levels which are taken out, however, have on average large velocities and therefore the average velocity of the remaining levels is different from zero. This effect can be removed by choosing a different unfolding mechanism which takes into account the fluctuations in the spectrum due to the bouncing ball orbits. This is done by defining

$$x_n = \bar{N}(k_n) + N_{bb}(k_n) , \qquad (93)$$

where $N_{bb}(k)$ is the semiclassical contribution of the bouncing ball orbits to the spectral staircase. We would like to note that this unfolding procedure alone already leads to a more symmetric velocity distribution and reduces slightly the peak of the curvature distribution at small curvatures.

We varied the number of levels that were removed and found that we had to neglect about 25 % of the spectrum until most of the non–generic contributions to the distributions were removed. The resulting velocity and curvature distributions, obtained by unfolding with (93), are shown in figures 10 and 11. The agreement with the random matrix results is now much better than before. One can still see some small deviations, but the main part of the non–generic contributions are removed. This indicates that wave functions that are scarred along the bouncing ball orbits are really the reason for the deviation from the random matrix results.

We have calculated the velocity and curvature distributions also for several other values of the parameter $\alpha$. Over a large range of this parameter the results are very similar to those for $\alpha = 0$. The velocity distributions are asymmetric and have a tail at large velocities which is larger than for a Gaussian and the curvature distributions show an excess of small curvatures. Also, if one removes again the part of the spectrum which is mostly affected by the bouncing ball orbits and performs the unfolding with eq. (93) the agreement with the random matrix results becomes quite good. However, as one approaches the value $\alpha = \pi/2$ the distributions become qualitatively different. In figure 12 the distribution of all velocities in the range $100 < k_n < 300$, unfolded without bouncing ball terms, is shown for $\alpha = \pi/2$. It is compared to a Gaussian with the same mean and standard deviation.

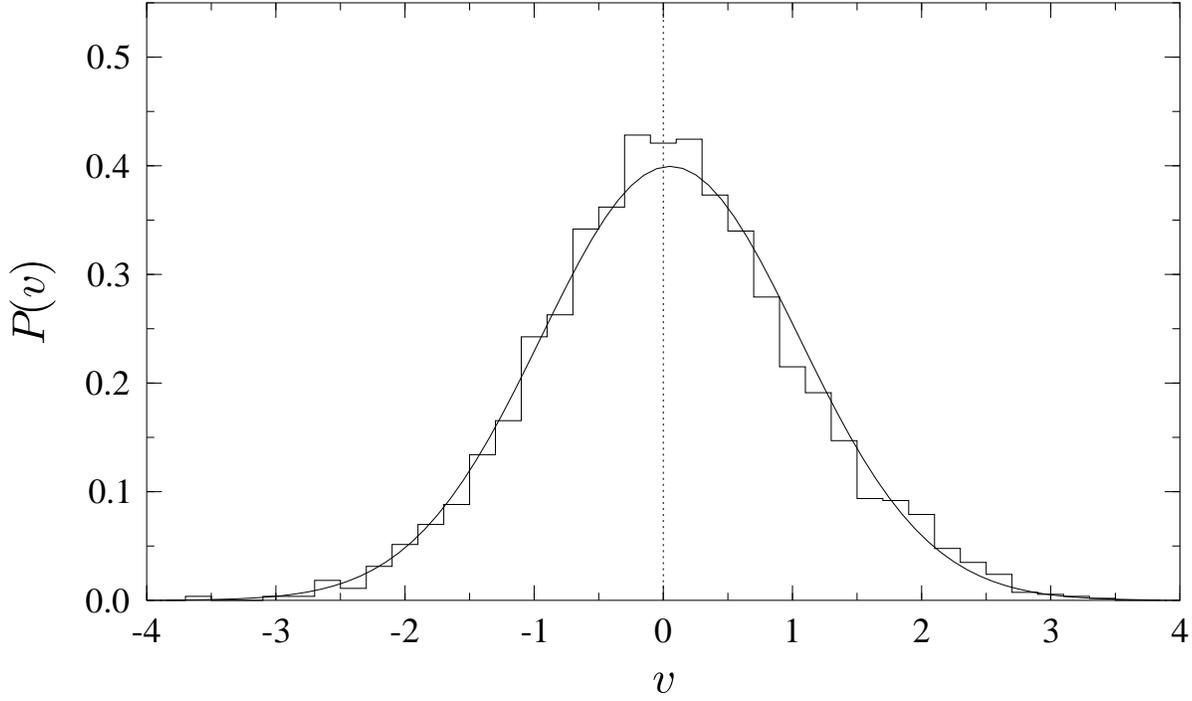

Figure 10: The velocity distribution $P(v)$ of a part of the spectrum at $\alpha = 0$ in comparison with a Gaussian, as described in the text.

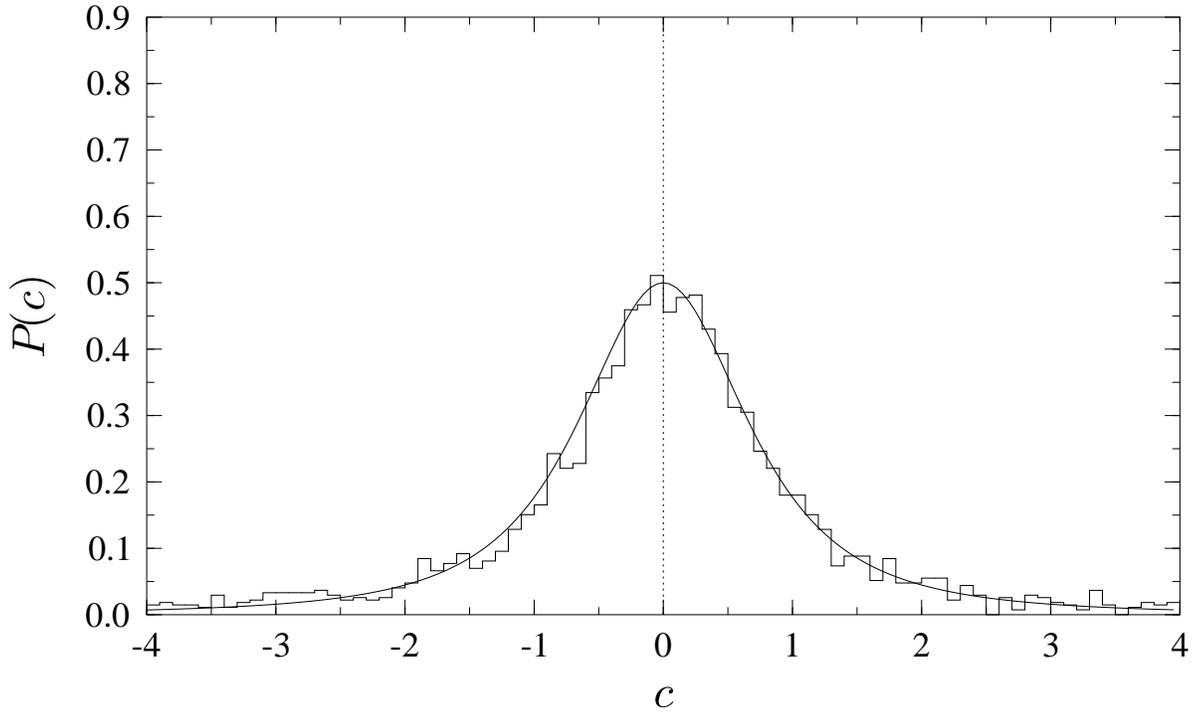

Figure 11: The curvature distribution $P(c)$ of a part of the spectrum at $\alpha = 0$ in comparison with the GOE-distribution of eq. (92), as described in the text.

Here the agreement with the Gaussian is better than for the distribution for $\alpha = 0$ in figure 8. The distribution is more symmetric, the peak at $v = 0$ is lower and there is a smaller excess of large velocities. The curvature distribution is shown in figure 13. It differs strongly from the $\alpha = 0$ case. There are much less contributions from small curvatures and the deviations from the random matrix curve are mainly due to an asymmetry in the distribution.

The different behavior near $\alpha = \pi/2$ can again be explained by the apparent lines in the spectrum. As was discussed in the last section, these lines are not quite horizontal near $\alpha = \pi/2$, but they have a negative slope and negative curvature. As a consequence, when the levels are unfolded they have smaller velocities and larger (negative) curvatures than in the case $\alpha = 0$. This is exactly the difference that is observed in the velocity and curvature distributions.

We also tried for $\alpha = \pi/2$ to improve the agreement with the random matrix results by removing a part of the spectrum and unfolding with eq. (93). However, in this case the agreement does not improve. The peak of the velocity distribution goes down, but the distribution is slightly asymmetric and the peak of the curvature distribution decreases below that of the random matrix result. Presently, we do not have a satisfactory explanation for this result. Possible explanations are that due to the stronger $\alpha$-dependence of the horizontal lines they have an influence also on more distant energy levels, or that the mixed boundary conditions do not behave generically near Neumann boundary conditions, at least in the considered energy range.

A definite interpretation of our results would also require an examination of wave functions. However, in case that there are many scarred wave functions at $\alpha = \pi/2$, our results seem to indicate, that scarred states do not necessarily have to lead to an accumulation of small curvatures. A prerequisite for such an accumulation is an almost linear change of the corresponding energy levels with the parameter $\alpha$, which is not necessarily the case.

Finally, we add a remark on the choice of the parameter $\alpha$. The curvature distribution is not independent on this choice if it is evaluated within a restricted energy range. We chose a parameter dependence that is symmetric with respect to Dirichlet and Neumann boundary conditions, but in principle, one could define a different parameter $\lambda = f(\alpha)$ and then the curvature distribution would change. For example we calculated at $\alpha = \pi/4$ the curvature distribution also for the $\kappa$-dependent boundary conditions of eq. (4) and we got a slightly different result than for the $\alpha$-dependence. (At $\alpha = \pi/2$ there is no difference, since the Taylor expansion of the sine has no quadratic term.) Examinations of the universality of the curvature distribution thus require also a careful definition of the parameter dependence.

# 6 Discussion

In this paper we have extended the semiclassical methods for treating billiard systems to situations where the boundary conditions interpolate between the standard Dirichlet and Neumann conditions. The parametric dependence on the boundary condition was shown to be a very useful tool in the spectral analysis and in the context of the parametric spectral statistics, the variation of the boundary condition is similar to the variation of any external parameter in a Hamiltonian system. It should be born in mind, however, that the boundary condition is strictly a wave (quantum) feature and as such, has no classical analogue. In this sense the system we consider here is different from the usual parameter dependent Hamiltonian systems, where the parameter has similar meanings in the quantum and classical representations. We shall address this point in the present section and show that, in one limit, there exists an alternative interpretation of the mixed boundary conditions. We will discuss also some results which can further illuminate some features of the semiclassical approximation for mixed boundary conditions.

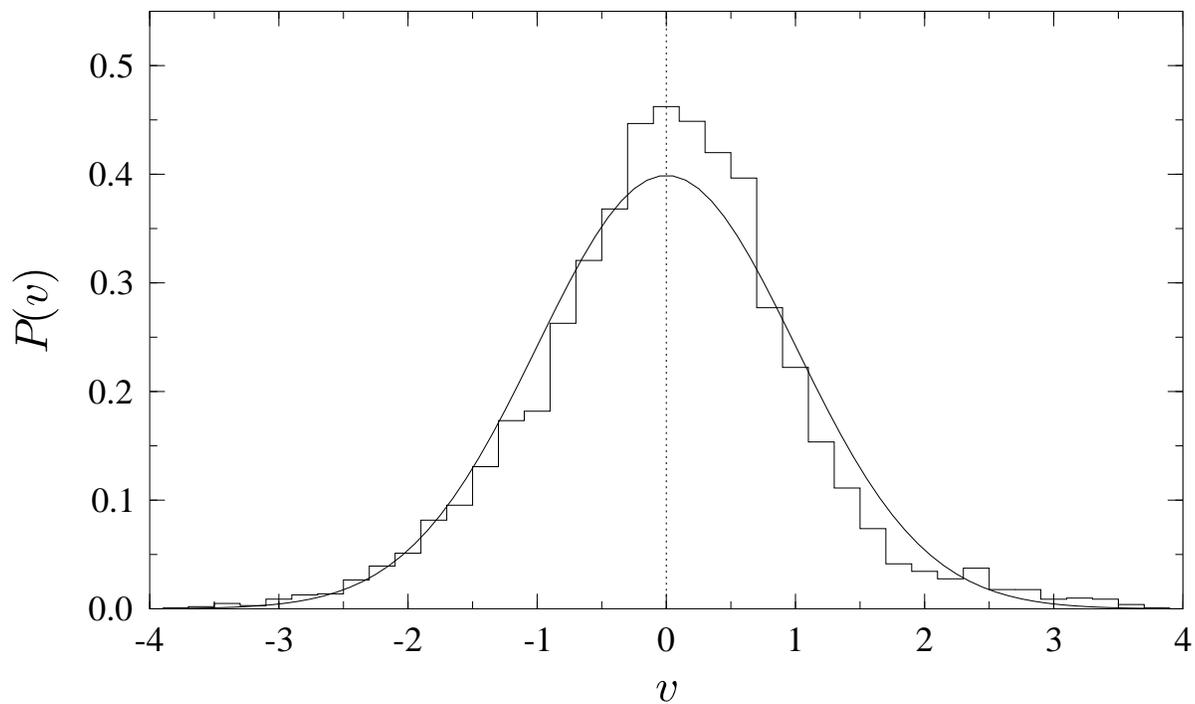

Figure 12: The velocity distribution $P(v)$ at $\alpha = \pi/2$ in comparison with a Gaussian.

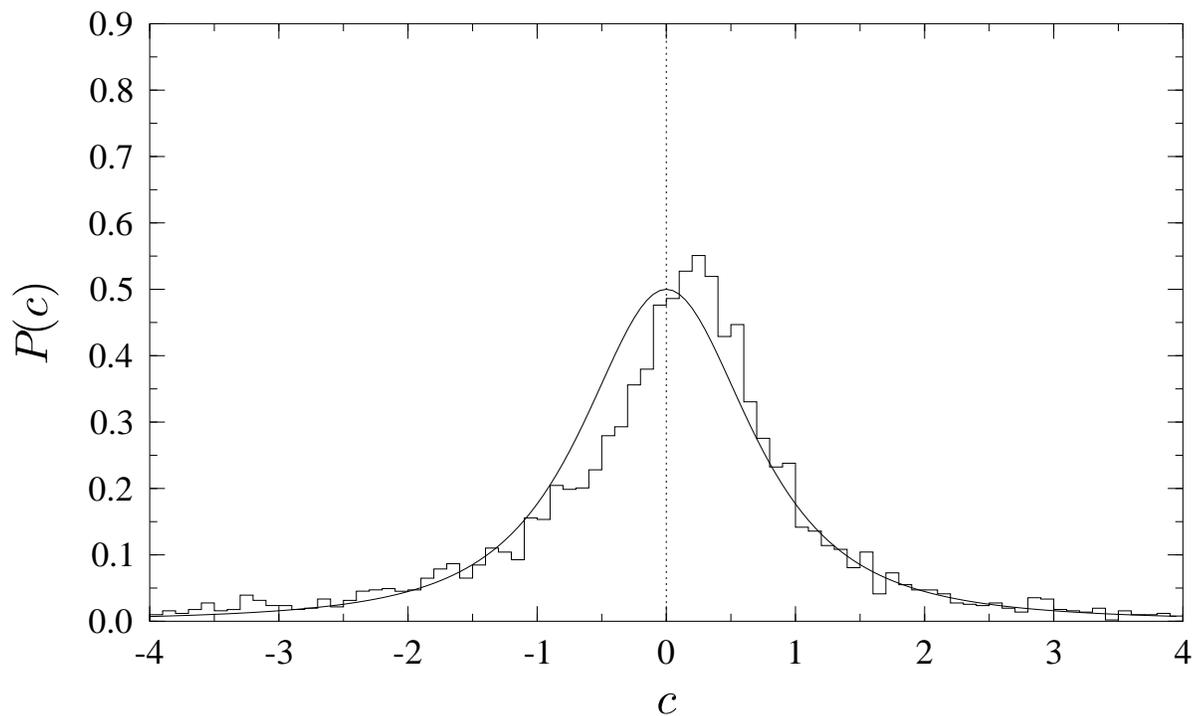

Figure 13: The curvature distribution $P(c)$ at $\alpha = \pi/2$ in comparison with the GOE-distribution of eq. (92).

We shall show first, that the variation of the spectral density in the vicinity of the Dirichlet boundary condition can be related to a variation of the boundary itself. The later certainly has a similar meaning in both the classical and the quantum pictures. We rewrite the mixed boundary conditions, eq. (4), as

$$\psi(\vec{r}) + \epsilon \partial_{\hat{n}} \psi(\vec{r}) = 0 \ , \ \vec{r} \in \Sigma \ , \tag{94}$$

where $\epsilon = \kappa^{-1}$, $\epsilon > 0$. This defines a Green function, $G(\vec{r}', \vec{r}''; \epsilon)$, and its derivative with respect to $\epsilon$ near $\epsilon = 0^+$ (Dirichlet case), $g(\vec{r}', \vec{r}'')$, can be easily shown to satisfy

$$g(\vec{r}', \vec{r}'') + \int_\Sigma ds(\vec{\rho}) \ G_0(\vec{r}', \vec{\rho}) \ \partial_{\hat{n}} g(\vec{\rho}, \vec{r}'') = -\int_\Sigma ds(\vec{\rho}) \ \partial_{\hat{n}} G_0(\vec{r}', \vec{\rho}) \ \partial_{\hat{n}} G(\vec{\rho}, \vec{r}''; \epsilon = 0) \ , \tag{95}$$

where $G_0(\vec{r}, \vec{r}')$ is the free Green function. Let us now define a uniform inflation of the boundary, which is affected by shifting each point $\vec{\rho}$ on the boundary a distance $\nu$ along the normal:

$$\Sigma(\nu): \ \vec{\rho} \to \vec{\rho} + \nu \hat{n}(\vec{\rho}) \ , \quad \vec{\rho} \in \Sigma \ . \tag{96}$$

Note, that this inflation is not just a rescaling of the coordinates, except for special cases like the circle billiard and regular polygons. Hence, the structure of the periodic orbits will in general change. Keeping the boundary condition to be Dirichlet ($\epsilon = 0$) for all values of $\nu$, one can show after some calculation, that $\partial G(\vec{r}', \vec{r}''; \nu)/\partial \nu|_{\nu=0}$ satisfies the same integral equation (95) as for $g(\vec{r}', \vec{r}'')$. Thus we establish the relation between Green functions

$$\left. \frac{\partial G(\vec{r}', \vec{r}''; \epsilon, \nu)}{\partial \epsilon} \right|_{\epsilon=\nu=0} = \left. \frac{\partial G(\vec{r}', \vec{r}''; \epsilon, \nu)}{\partial \nu} \right|_{\epsilon=\nu=0} . \tag{97}$$

To derive the relation between the spectral densities we use for $k > 0$

$$d(k; \epsilon, \nu) = -\frac{2k}{\pi} \lim_{\delta \to 0^+} \int_A d\vec{r} \, \text{Im} \, G(\vec{r}, \vec{r}; k^2 + i\delta; \epsilon, \nu) \ . \tag{98}$$

The derivative with respect to $\epsilon$ affects only the integrand, while the derivative with respect to $\nu$ affects also the integration domain. However, since we consider Dirichlet boundary conditions ($\epsilon = 0$), we have $G(\vec{r}, \vec{r}; \epsilon = 0) = 0$ for $\vec{r} \in \Sigma$, and the second contribution is 0. Therefore we conclude:

$$\left. \frac{\partial d(k; \epsilon, \nu)}{\partial \epsilon} \right|_{\epsilon=\nu=0} = \left. \frac{\partial d(k; \epsilon, \nu)}{\partial \nu} \right|_{\epsilon=\nu=0} . \tag{99}$$

This exact result becomes intuitively clear in the semiclassical limit: For the smooth part of the level density we can use $\partial A/\partial \nu|_{\nu=0} = L$, where $A, L$ are the billiards area and perimeter, respectively. When this is inserted into eq. (7) we get to leading order:

$$\left. \frac{\partial \bar{d}(k; \epsilon, \nu)}{\partial \epsilon} \right|_{\epsilon=\nu=0} = \left. \frac{\partial \bar{d}(k; \epsilon, \nu)}{\partial \nu} \right|_{\epsilon=\nu=0} = \frac{Lk}{2\pi} \ . \tag{100}$$

(For smooth billiards one gets also the next term by similar methods.) As for $d_{osc}$, consider only the contributions of generic periodic orbits which are far from tangency to the boundary. In the Gutzwiller sum, only the fast varying derivatives of the actions $kL_\gamma$ with respect to $\nu$ will be important in the semiclassical limit. Simple geometry shows that $\partial L_\gamma / \partial \nu|_{\nu=0} = 2 \sum_{i=1}^{n_\gamma} \cos \theta_\gamma^i$, and together with the relation $\partial/\partial\epsilon|_{\epsilon=0} = b\, \partial/\partial\alpha|_{\alpha=0}$ one gets the desired equivalence (see eq. (86)).

Even though the boundary condition is a pure quantum feature, its expression in the semiclassical spectral density depends on the underlying classical dynamics and deserves some more discussion. In particular, we shall study now the phase factor which multiplies the contributions of the generic unstable periodic orbits (9). The main feature of this phase is that for the mixed boundary conditions it depends on both, $k$ and the angles of incidence $\theta_i, i = 1, \ldots n$ along the periodic orbit. This should be contrasted with the Dirichlet and Neumann boundary conditions, where a reflection leads to a simple factor of $(-1)$ or $(+1)$, respectively.

Let us consider the phase factor for very long periodic orbits. We are interested in orbits that fill the phase space uniformly, so that the phase factor can be calculated approximately by an ergodic averaging over phase space. The appropriate coordinates for doing this are the length along the perimeter $s$ and the canonical momentum $p = \cos\varphi$, where $\varphi$ is the angle of the reflected trajectory with the (directed) tangent. The phase $\Phi$ is then obtained as

$$\begin{aligned}
\Phi &= 2 \sum_{i=1}^{n} \arctan\left(\frac{k}{\kappa} \cos\theta_i\right) \\
&= \frac{n}{L} \int_0^L ds \int_{-1}^1 dp \, \arctan\left(\frac{k}{\kappa} \sqrt{1-p^2}\right) \\
&= n\pi \left( \sqrt{1 + \left(\frac{\kappa}{k}\right)^2} - \frac{\kappa}{k} \right) \, .
\end{aligned} \quad (101)$$

The number of reflections on the boundary $n$ can be expressed by $n = l/<c>$, where $l$ is the length of the orbit and $<c>$ is the mean chord length, i.e. the mean distance between two consecutive reflections on the boundary. Denoting by $c(s,\varphi)$ the length of a chord that starts at $s$ with an angle $\varphi$ with respect to the tangent, the mean chord length is obtained by

$$\begin{aligned}
<c> &= \frac{1}{2L} \int_0^L ds \int_0^\pi d\varphi \, \sin\varphi \, c(s,\varphi) \\
&= \frac{1}{2L} \int_0^{2\pi} d\alpha \int_0^L ds \, \sin\varphi(\alpha) \, c(s,\varphi(\alpha)) \\
&= \frac{\pi A}{L} \, ,
\end{aligned} \quad (102)$$

where a transformation to an angle $\alpha$ with respect to a fixed direction in space has been made and $c(s,\varphi) = 0$ for directions that point outside of the billiard. The integral over $ds$ gave the area $A$, since $ds \cdot \sin\varphi$ is the component of $ds$ orthogonal to the chord. With this result the phase $\Phi$ is given by

$$\Phi = l \frac{L}{A} \left( \sqrt{1 + \left(\frac{\kappa}{k}\right)^2} - \frac{\kappa}{k} \right) \, . \quad (103)$$

This additional phase can be interpreted as a small correction to the classical action $l \cdot k$ which is the leading term in the semiclassical phase. The expression which multiplies $l$ in eq. (103) leads to a mean shift in the spectrum with respect to the Dirichlet value:

$$k_n - k_n^D = -\frac{L}{A}\left(\sqrt{1 + \left(\frac{\kappa}{k_n}\right)^2} - \frac{\kappa}{k_n}\right) . \qquad (104)$$

On the other hand the mean spectral shift due to a change in the mean number of levels is semiclassically given to leading order by (7)

$$k_n - k_n^D = -[\bar{N}(k_n) - \bar{N}^D(k_n)]\overline{\Delta k} = -\frac{L}{A}\left(\sqrt{1 + \left(\frac{\kappa}{k}\right)^2} - \frac{\kappa}{k_n}\right) , \qquad (105)$$

where $\overline{\Delta k} = 1/\bar{d}(k)$ is the mean distance between wave numbers. The result is identical to eq. (104). It shows that the ergodic average over the additional phase is responsible for the mean shift of the energy levels with respect to the Dirichlet case. The deviations of the phase factors of periodic orbits from the ergodic mean thus lead to fluctuations of the energy levels around their mean shift. The fact that the mean shift in the energy levels, which is derived from considering very short orbits, can also be obtained from the ergodic average of very long orbits is another example of the bootstrapping property in periodic–orbit theory [36].

The last comment is a word of caution. The mixed boundary conditions were shown to appear in the expressions for the smooth and the oscillatory components of the spectral density. The theory presented here for the smooth part gives the first few leading terms in the asymptotic expansion of $\bar{d}(k)$ in powers of $k^{-1}$. This is not true for the oscillatory contribution, where only the leading contribution was considered.

# 7 Acknowledgments

We are obliged to Dr. E. Bogomolny for interesting discussions and suggestions which were very helpful during various stages of the work. M. S. would like to thank Prof. H. J. Stöckmann for helpful discussions. The research reported here was supported by the Minerva Center for Nonlinear Physics of Complex Systems and by the US–Israel Binational Science Foundation (BSF). M. S. wishes to acknowledge financial support by the MINERVA, the Einstein Center and the Alexander von Humboldt-Stiftung.

# A  The Smooth Part of the Spectral Density for the Semi-circle Billiard

In this appendix, the smooth part of the spectral staircase $\bar{N}(k)$ is derived for a semi-circle with mixed boundary conditions on the circular part of the boundary and Dirichlet or Neumann boundary conditions on the diameter, respectively. The derivation is done by using the methods of section 2.1. From the results the contribution of a 90° corner is extracted.

The Green function for the semi-circle billiard is obtained from that of the circle billiard by the method of images:

$$\tilde{G}(\vec{r}, \vec{r}', s^2) = \tilde{G}^c(\vec{r}, \vec{r}', s^2) \mp \tilde{G}^c(\vec{r}, \hat{R}\vec{r}', s^2) , \qquad (106)$$

where the upper and lower sign correspond to Dirichlet and Neumann boundary conditions on the diameter, respectively and the operator $\hat{R}$ denotes the reflection on the line $y = 0$. The upper index $c$ will from now on denote the quantities for the full circle billiard. From eq. (106) the Green function for the semi-circle follows as

$$\tilde{G}(\vec{r}, \vec{r}', s^2) = -\frac{1}{2\pi} \sum_{n=-\infty}^{\infty} I_n(sr_<)[K_n(sr_>) + a_n I_n(sr_>)] \left[\cos n(\theta - \theta') \mp \cos n(\theta + \theta')\right] , \qquad (107)$$

Subtracting the free Green function and taking the trace results in

$$\begin{aligned}\tilde{K}(s^2) &= \int_0^R \mathrm{d}r\, r \int_0^\pi \mathrm{d}\theta\, [\tilde{G}(\vec{r}, \vec{r}', s^2) - \tilde{G}_0(\vec{r}, \vec{r}', s^2)] \\ &= \frac{1}{2}\tilde{K}^c(s^2) \mp \frac{R^2}{4}[f(0,s) - I_0(sR)K_0(sR) + I_0'(sR)K_0'(sR)] . \end{aligned} \qquad (108)$$

Now, the asymptotic expressions for the Bessel functions for large arguments are inserted and after expanding in powers of $1/(sR)$ one obtains

$$\tilde{K}(s^2) = \frac{1}{2}\tilde{K}^c(s^2) \mp \frac{R^2}{4}\left[\frac{\kappa/s - 1}{2(sR)^2(\kappa/s + 1)} - \frac{1}{sR}\right] . \qquad (109)$$

From this the contributions to the mean level density and to the mean spectral staircase follow as

$$\bar{d}(k) = \frac{1}{2}\bar{d}^c(k) \mp \frac{R}{2\pi} \mp \frac{1}{2\pi k}\frac{\frac{\kappa}{k}}{1 + \left(\frac{\kappa}{k}\right)^2} \qquad (110)$$

and

$$\bar{N}(k) = \frac{1}{2}\bar{N}^c(k) \mp \frac{kR}{2\pi} \mp \left(\frac{1}{2\pi}\arctan\frac{k}{\kappa} - \frac{1}{8}\right) , \qquad (111)$$

where the constant term has been determined as discussed in section 2.1. The term $\mp kR/(2\pi)$ in eq. (111) is the contribution of the diameter to $\bar{N}(k)$. The last term in eq. (111) is the contribution of the two corners and thus it follows that the contribution of a 90° corner with mixed boundary conditions on one side and Neumann and Dirichlet boundary conditions, respectively, on the other side is given by

$$\mp \left(\frac{1}{4\pi}\arctan\frac{k}{\kappa} - \frac{1}{16}\right) . \qquad (112)$$